\documentclass[12pt,a4paper]{article}
\usepackage{amsmath}
\usepackage{amssymb}
\usepackage{amsthm}
\usepackage{float}
\usepackage{amsfonts}
\usepackage{graphicx}
\usepackage{verbatim}
\usepackage[left=2cm,right=2cm,top=3cm,bottom=2.5cm]{geometry}
\usepackage[numbers]{natbib}
\usepackage[utf8]{inputenc}
\usepackage[usenames,dvipsnames,svgnames]{xcolor}
\usepackage[colorlinks=true,
      linkcolor=red,
      urlcolor=gray,
      citecolor=blue]{hyperref}

\def\myalign#1{%
  \def\trule{\noalign{\smallskip\hrule\medskip}}
  \def\nebc{\nearrow\bigcup}
  \def\sebc{\searrow\bigcup}
  \def\pminf{{}_{-\infty}|^{+\infty}}
  \let\Inf\infty
  \def\amp{&} 
  \vbox{\mathsurround0pt\openup1\jot
    \halign{%
      &$\displaystyle##\hfil\tabskip0pt$&\amp##\tabskip1em\crcr
      \noalign{\hrule height1pt\smallskip}#1\noalign{\smallskip\hrule height1pt}\crcr}}}
      
\begin{document}
\begin{center}
\textbf{On inflationary parameters in scalar-tensor theories}
\end{center}
\hfill\\
Beatrice Murorunkwere$^{1}$, Fidele Twagirayezu $^{1}$, Albert Munyeshyaka$^{2}$, Joseph Ntahompagaze$^{1}$ and Abraham Ayirwanda$^{1}$\\
\hfill\\
$^{1}$Department of Physics, College of Science and Technology, University of Rwanda, Rwanda\;\;\; \; \;\hfill\\ 
$^{2}$Department of Physics, Mbarara University of Science and Technology, Mbarara, Uganda\\
\hfill\\
Correspondence:munalph@gmail.com\;\;\;\;\;\;\;\;\;\;\;\;\;\;\;\;\;\;\;\;\;\;\;\;\;\;\;\;\;\;\;\;\;\;\;\;\;\;\;\;\;\;\;\;\;\;\;\;\;\;\;\;\;\;\;\;\;\;\;\;\;\;\;\;\;\;\;\;\;\;\;\;\;\;\;
\begin{center}
\textbf{Abstract}
\end{center}
The equivalence between $f(R)$ and scalar-tensor theories is revisited, we consequently  explored different  $f(R)$ models. After consideration of specific  definition of the scalar field, we  derived the  potentials $V(\phi)$ for each $f(R)$ model focusing on the early Universe, mostly the inflation epoch.  For a given potential, we applied  the slow-roll approximation approach  to each $f(R)$ model and  obtained the expressions for the spectral index $n_{s}$ and tensor-to-scalar ratio $r$. We determined the corresponding numerical values  associated with each  of the $f(R)$ models. Our results showed that for certain choice of  parameter space, the values of $n_{s}$ and $r$ are consistent with the Planck survey results  and others produce numerical values  that are in the same range as suggested by Planck data. We further constructed the Klein-Gordon equations $(KGE)$ of each $f(R)$ model. We found numerical solutions to  each KGE considering different values of free parameters and initial conditions of each $f(R)$ model. All models showed that the scalar field decreases as time increases, an indication that there is less content of the scalar field in the late Universe.
\\
\hfill\\
\textit{keywords:} $f(R)$ gravity --- scalar-tensor theory --- inflation ---scalar fieldThis manuscript was accepted for publication in International Journal of Geometric Methods in Modern Physics, doi: 10.1142/S0219887822501456

\section{Introduction}\label{INTRODUCTION}
The observations showed that the Universe has undergone a couple of phases of cosmic  acceleration (early-time and late-time acceleration).  The early-time acceleration happened at the early stages of the evolution of  Universe and is known as inflation. During this phase, scalar field was dominating over standard matter and it was first introduced to solve a number of problems in the standard cosmology  \cite{guth1981cosmological,linde2008inflationary}. During the inflationary era, the Universe increased its size at an exponential rate so it expands quite quickly in a short time. To explain inflationary cosmology, many models were proposed, however, Planck data \cite{ade2016planck}  excluded  some from the list of viable inflationary models  \cite{martin2016have}. Some of these models use scalar field as good candidates to drive inflation and explain how to relate theoretical predictions to observable quantities \cite{gorbunov2011introduction} while others use modified theories of gravity \cite{nojiri2007econf,capozziello2011extended,nojiri2017modified} for such purpose. The $f(R)$ theory models are among the modified  gravity theories which remain viable after the release of Planck data \cite{odintsov2016inflationary,odintsov2017inverse}.

There is also late-time accelerating phase as confirmed by datasets coming from different sources, such as temperature fluctuations of the Cosmic Microwave Background (CMB) radiations, large-scale distribution of galaxies and supernovae surveys
 \cite{riess1998observational,perlmutter1999constraining} . These two phases can not be achieved, if one considers  General Relativity (GR) without the introduction of some additional fields. One way to explain the acceleration of the Universe is by using modified gravity in which  the gravitational theory is modified compared to $GR$. One of the modifications to $GR$ is the $f(R)$ theories of gravity where $R$ is the Ricci scalar  \cite{capozziello2002curvature,carloni2005cosmological,cembranos2006dark}. In the $f(R)$ gravity, the action for the gravitational interaction is written as a generic analytic function of the Ricci scalar \cite{nojiri2006dark}. The studies of $f(R)$ theories have attracted many attention due to the fact that it has been proven in different ways that they can produce late-time acceleration  without introduction of dark energy  and cosmological constant \cite{sotiriou2006nearly,sawicki2007stability}.
 
Different studies have been done for different reasons using $f(R)$ models. For example, investigations related to the cosmological viability  of perturbations of class of $f(R)$ theory of gravity in assisting large-scale structure formation were done in \cite{carloni2010covariant}. In addition to that, Sotiriou \cite{sotiriou2006f} investigated the relationship between $f(R)$ and scalar-tensor $(ST)$ theories.  Among other applications, $f(R)$  theory of gravity offers the possibility of unified description of the early-time and late-time acceleration \cite{capozziello2000nonminimal,nojiri2003modified,capozziello2006unified,appleby2008aspects,de2011generalizing,sepehri2015unifying,odintsov2019unification,yashiki2020local,odintsov2020geometric,odintsov2020aspects,odintsov2021r,oikonomou2021rescaled,oikonomou2021unifying}. Furthermore, in \cite{starobinsky1980new}, cosmological inflation (one of the early Universe epochs) in modified gravity  was studied first by starobinsky and showed that $f(R)$ corrections to the standard General Relativity $GR$ can lead to an early phase of the site expansion  and several other studies have been conducted since then \cite{nojiri2003modified, nojiri2007unifying, nojiri2008future, nojiri2017modified, barrow2018reconstructions,munyeshyaka2021cosmological}. In this study, we will focus on the  $f(R)$ gravity description  of inflation connected with ST theory.\\
 
Computing the inflationary parameters in $f(R)$ theory of gravity  has been studied extensively  over decades using different approaches.   For instance, in \cite{ntahompagaze2017f}, they applied slow-roll approximation method for each of the four $f(R)$ models and computed the expressions of the spectral index $n_{s}$ and tensor-to-scalar ratio $r$  considering the scalar field $\phi=f'-1$. The numerical values of $n_{s}$ and $r$ obtained were very close to the ones from Planck data. The same authors computed two inflationary parameters spectral index $n_{s}$ and tensor-to scalar ratio $r$  for $f(R)$ models based on the definition of the scalar field $\phi=f'$. The produced values of $n_{s}$ were in the same range as the values suggested from the observations but for $r$, some of the $f(R)$ models suffered to produce values which are in agreement with observation;  \cite{ntahompagaze2018inflation}. 
The exploration of how the Einstein’s frame can be used to reconstruct $f(R)$ models by specification of the potential has been pointed out, many $f(R)$ models have been analysed and values of $n_{s}$ and $r$ were computed for a given number of e-fold $N$ in  \cite{sebastiani2014nearly}.\\

In \cite{oikonomou2015singular,myrzakulov2016inflationary,oikonomou2021refined}, the computation of both $n_{s}$ and $r$ was made with  the consideration of higher-derivative quantum gravity that contains Gauss– Bonnet terms, the obtained values of $n_{s}$ and $r$  are in agreement with the observations. It was also shown in \cite{odintsov2017unification} using  constant-roll approximations for both models that contain logarithmic and exponential terms of $f(R)$ gravity that the values of $n_{s}$ and $r$ produced  are in the range predicted by the observations.\\
Over  recent years, much progress has been made on inflationary parameters, however, there seems to be one definition of the scalar field $\phi=ln(f')$ that has not been explored for the same purpose. In the current paper, we considered this way of defining the scalar field based on our previous work \cite{murorunkwere20211+} and compute two inflationary parameters (spectral index $n_{s}$ and tensor-to- scalar ratio $r$) and compare them with Planck survey results.\\
 
The outline of the paper is as follows:
In Section {\color{red}2}, we gave a brief overview of the actions and field equations.
In Section {\color{red}3}, we derived the expressions of the potential, inflation parameters and Klein-Gordon equations for each  $f(R)$ pedagogical model.
In Section {\color{red}4}, we specialised the discussion to  each toy of $f(R)$ models.
In Section {\color{red}5}, we present our conclusion.\\
The adopted spacetime signature is (-+++) and unless stated otherwise, we have used  the reduced Planck units, $\hbar= 8\pi G=c=1$ where $\hbar$ is Planck’s constant (reduced), $G$ is the gravitational constant and $c$ is the speed of light.
The symbol $ \nabla$ refers to covariant derivative, $ \partial$ is partial differentiation and the over dot shows differentiation with respect to cosmic time.

\section{Field equations}
One of the most widely explored alternatives to $GR$ in the context of late-time acceleration of the Universe are $f(R)$ theories of gravity. These models are generally obtained by including higher order curvature invariants in the Einstein-Hilbert action.
The general action of $f(R)$ gravity theory is given by \cite{sotiriou2010f,clifton2012modified} 
\begin{equation}
A_{f(R)}=\frac{1}{2 \kappa} \int d^{4} x \sqrt{-g} \Big.\Big[f(R)+2\mathcal{L}_{m}(g_{\mu\nu},\psi)\Big.\Big],
\end{equation}
where $ \kappa=8\pi G$ is the Einstein gravitational constant, $R$ is the curvature scalar and $\mathcal{L}$ is the matter Lagrangian .\\
The two variation principles that one can apply to action in order to derive Einstein-field equations are:  the metric and Palatini variations and this results in metric or Palatine $f(R)$ gravity, according to which variational principle is applied.
Varying the action  in Eqn.{\color{blue}(1)} with respect to the metric  results in metric formalism and produces the following field equation
\begin{equation}
G_{\mu\nu}=\frac{1}{f'}\Big.\Big[T^{m}_{\mu\nu}+\frac{1}{2}g_{\mu\nu}\Big.\Big(f-Rf'\Big.\Big)+ \nabla_{\mathbf{\mu}}  \nabla^{\mathbf{\mu}}f' - g_{\mu\nu} \nabla_{\mathbf{\alpha}}  \nabla^{\mathbf{\alpha}} f' \Big.\Big],
\end{equation}
where $f=f(R)$, $f'$ denotes the derivative of the function $f$ with respect to its argument $R$, $  \nabla_{\mathbf{\alpha}} \nabla^{\mathbf{\alpha}} $ is the notation for covariant D'Alembert operator and 
$T^{m}_{\mu\nu}=-\frac{2}{\sqrt{-g}}\frac{\partial (\sqrt{-g}\mathcal{L}_{m})}{\partial g^{\mu\nu}}$,
is the energy momentum-tensor of the standard matter.\\
The general action that represents the scalar-tensor theory is given by \cite{sotiriou2006f}
\begin{equation}
 A_{ST}=\frac{1}{2 \kappa}\int d^4x \sqrt{-g} \Bigg[ \Bigg.\frac{y(\phi)}{2}R-\frac{W(\phi)}{2}\nabla_{\mu}\phi \nabla^{\mu}\phi-U(\phi)+2\mathcal{L}_{m}(g_{\mu\nu},\psi)\Bigg. \Bigg],
\end{equation}
where $ U(\phi)$ is the potential of the scalar field $\phi$, $ y(\phi)$ and $ W(\phi) $ are some function of $ \phi$.
Setting $y(\phi)=\frac{\phi}{ \kappa}$, $ W(\phi)=\frac{W}{ \kappa\phi}$ and $ U(\phi)=\frac{V(\phi)}{\phi}$, the scalar-tensor theory reduces to Brans-Dicke theory of gravity as
\begin{equation}
 A_{BD}=\frac{1}{2 \kappa} \int d^{4}x \sqrt{-g}\Big[ \Big.\phi R-\frac{W}{\phi} \nabla_{\mu}\phi \nabla^{\mu}\phi-V(\phi)+2\mathcal{L}_{m}(g_{\mu\nu},\psi)\Big. \Big].
\end{equation}
For vanishing coupling parameter, $ W=0$, the Brans-Dicke theory reduces to 
\begin{equation}
 A_{BD}=\frac{1}{2 \kappa}\int d^{4}x \sqrt{-g}\Big[ \Big.\phi R-V(\phi)+2\mathcal{L}_{m}(g_{\mu\nu},\psi)\Big. \Big].
\end{equation}
For a given $f(R)$ Lagrangian, one can define an auxiliarly field $ \chi$ such that it is a function of the scalar field $ \phi$ as $ \chi(\phi)$ so that the action of $f(R)$ gravity given by Eqn. {\color{blue}(1)} becomes
\begin{equation}
 A_{f(R)}=\frac{1}{2 \kappa} \int \sqrt{-g}\Big[ \Big.f(\chi)+f'(\chi)(R-\chi)+2\mathcal{L}_{m}(g_{\mu\nu},\psi)\Big. \Big],
\end{equation}
where variation with respect to $ \chi$ gives
$ f''(\chi)(R-\chi)=0$, if $ f''(\chi)=0$,
therefore $ \chi=R$ if $ f''\neq 0$.
Setting the potential as
\begin{equation}
 V(\phi)=\chi(\phi)\phi-f(\chi(\phi)),
\end{equation}
the action of $f(R)$ gravity given by Eqn. {\color{blue}(6)} takes the form
\begin{equation}
 A_{f(R)}=\frac{1}{2\kappa}\int d^{4}x \sqrt{-g}\Big[ \Big.\phi R-V(\phi)+2\mathcal{L}_{m}(g_{\mu\nu},\psi)\Big. \Big].
\end{equation}
The action given by Eqn. {\color{blue}(8)} is equal to the action of Brans-Dicke theory for vanishing coupling parameter given by  Eqn. {\color{blue}(5}),
hence $f(R)$ is a special case of the scalar-tensor theory.\\
The scalar field $\phi$ obeys the Klein-Gordon equation as \cite{frolov2008singularity}
\begin{equation}
\square \phi-\frac{1}{3}\Big.\Big[2f-e^{\phi} R+T^{m} \Big.\Big]=0.
\end{equation} 
The energy-momentum tensor by definition is given by
\begin{equation}
T^{m}=\frac{\rho_{m}}{f'}\Big.\Big(3w-1\Big.\Big),
\end{equation}
where $w$ is the equation of state parameter. Using $ \square=-\frac{\partial^{2}}{\partial t^{2}}+\nabla^{2} $, we have the Klein-Gordon equation {\color{blue}(9)}  as
\begin{equation}
-\frac{\partial^{2}\phi}{\partial t^{2}}+\nabla^{2}\phi-\frac{1}{3}\Big.\Big[2f-e^{\phi}R+\frac{\rho_{m}}{f'}\Big.\Big(3w-1\Big.\Big)\Big.\Big]=0.
\end{equation}
Focusing to early Universe, the matter energy density $\rho_{m}$ is neglected over the scalar field and as results, the term $T^{m}$ in Eqn.{\color{blue}(11)} vanishes.
Also, the fact that the scalar field $\phi$ is assumed to be only time independent, we drop out the spatial dependence on the covariant  d'Alembert operator in Eqn.{\color{blue}(11)} . Thus, we write the KGE {\color{blue}(11)} as
\begin{equation}
-\frac{\partial^{2} \phi}{\partial t^{2}}-\frac{1}{3}\Big.\Big[2f-e^{\phi}R\Big.\Big]=0.
\end{equation}
The effective scalar field potential $V(\phi)$ is determined by \cite{frolov2008singularity}
\begin{equation}
V'(\phi)=\frac{dV}{d\phi}=\frac{1}{3}\Big.\Big(2f-\phi R\Big.\Big).
\end{equation}
Here and onwards,  prime indicates differentiation with respect to scalar field $\phi$.
In this study, we focus on the inflationary epoch and they are two approaches of treating the behaviour  of the inflation (slow-roll  and constant-roll approximations). The constant-roll inflation description, is an alternative approach to the standard slow-roll inflationary era, and its implications have recently been studied in the context of scalar-tensor theories \cite{odintsov2017inflation} and also in the context of $f(R)$  \cite{oikonomou2017reheating}. The slow-roll approximation approach is based on the fact that, during this phase, the scalar field was evolving slowly compared to potential.They are two different version of the slow-roll approximation. The first one \cite{copeland1993reconstructing} places conditions on the evolution of the Hubble parameter during the inflation and it is called Hubble slow-roll approximation (H-SR). The second one \cite{liddle1992cobe} places restrictions on the form of potential and requires the evolution of the scalar field to have reached its asymptotic form. This approach is most appropriate when studying  inflation in a specific potential and it is called potential slow-roll approximation (V-SR).  Based on the slow-roll approximation, the set of parameters (spectral index $n_{s}$ and tensor-to-scalar ratio $r$) that allows making connection with observations can be computed.\\
When provided with a potential from which the inflationary model is constructed, the slow-roll approximation approach  requires the smallness of two parameters  (both function of scalar field $\phi$) and they are defined as
 \begin{equation}
\varepsilon({\phi})=\frac{1}{2 \kappa^{2}} \Big.\Big(\frac{V'(\phi)}{V(\phi)}\Big.\Big)^{2},
\end{equation}
\begin{equation}
\eta(\phi)=\frac{1}{ \kappa^{2}}\Big.\Big(\frac{V''(\phi)}{v(\phi)}\Big.\Big).
\end{equation}
We refer to them as  potential slow-roll $(V-SR)$ parameters  \cite{komatsu2002pursuit}. The $\varepsilon(\phi)$ measures the slope of the potential and $\eta(\phi)$ measures the curvature.
The slow-roll approximation approach has been used in studying the dynamics of inflation and has the following two conditions \cite{liddle1994formalizing}:
\begin{itemize}
 \item The square of the time derivative of the slow-rolling scalar field has to be smaller than the slow-rolling scalar field potential, that is to say
 \begin{equation}
\Big.\Big( \frac{d\phi}{dt}\Big.\Big)^{2}<V(\phi).
 \end{equation}
 \item The second-order time derivative of slow-rolling scalar field is smaller than the derivative of the potential with respect to the scalar field $\phi$, that is
 \begin{equation}
 2\Big.\Big(\frac{d^{2}\phi}{dt^{2}}\Big.\Big)< \left|V'(\phi)\right| ,
 \end{equation}
 where prime indicates differentiation with respect to the scalar field $\phi$.
\end{itemize}
For slow-roll inflation, the spectral index $n_{s}$  and  the tensor-to-scalar ratio $r$ are respectively defined as \cite{bassett2006inflation,huang2014polynomial}
\begin{equation}
n_{s}=1-16 \varepsilon+2 \eta.
\end{equation}
and
\begin{equation}
r=16  \varepsilon.
\end{equation}
While dealing with $f(R)$ forms, various functionals were proposed including $R$-power function ($ f= R^{n}$), polynomial, exponential and logarithmic functions. Using each toy of $f(R)$ model, we get the relationship between scalar field and the Ricci scalar $R$. From there, we get the derivative of the effective potential $V'(\phi)$, the second derivative of the potential $V''(\phi)$  and the corresponding potential $V(\phi)$ in order to obtain spectral index $n_{s}$ and tensor-to scalar ratio $r$.
\section{The $f(R)$ models}
\subsection{Model1: $f(R)=\beta R^{n}$}
In $R^{n}$ gravity, the function $f$ is specified by a generic power of the Ricci scalar \cite{barrow1983stability}  as
\begin{equation}
 f(R)=\beta R^{n},
\end{equation}
where $\beta$ is a coupling constant such that $\beta=1$ for $GR$ case when $n=1$. This is the simplest and important  model when describing the early cosmological inflation and most widely studied form of $f(R)$ theory of gravity. Using the definition of the scalar field $\phi=ln(f')$ where $f'=\frac{df}{dR}$, Eqn.{\color{blue}(20)} in terms of scalar field becomes
\begin{equation}
f(\phi)=\beta \Big.\Big(\frac{1}{n\beta}e^{\phi}\Big.\Big)^{\frac{n}{n-1}}.
\end{equation}
Then by using Eqn.{\color{blue}(13)}, one can find an expression of  the derivative of the potential $V'(\phi)$ with respect to scalar field  as
 \begin{equation}
V'(\phi)=\frac{2}{3}\beta\Big.\Big(\frac{e^{\phi}}{n\beta}\Big.\Big)^{\frac{n}{n-1}}-\frac{1}{3}\phi(\frac{e^{\phi}}{n\beta})^{\frac{1}{n-1}}.
\end{equation}
The second derivative of the potential  of  Eqn.{\color{blue}(22)} yields
\begin{equation}
V''(\phi)=\frac{2\beta n}{3(n-1)}\Big.\Big(\frac{e^{\phi}}{\beta n}\Big.\Big)^{\frac{n}{n-1}}-\frac{\phi}{3(n-1)}\Big.\Big(\frac{e^{\phi}}{\beta}\Big.\Big)^{\frac{1}{n-1}}-\frac{1}{3}\Big.\Big(\frac{e^{\phi}}{\beta n}\Big.\Big)^{\frac{1}{n-1}}.
\end{equation}
The integration of the Eqn.{\color{blue} (22)} with respect to $\phi$, yields
\begin{equation}
V(\phi)=\frac{2\beta(n-1)}{3n}\Big.\Big(\frac{e^{\phi}}{\beta n}\Big.\Big)^{\frac{n}{n-1}}+\frac{(n-1)(n-1+\phi)}{3}\Big.\Big(\frac{e^{\phi}}{\beta n}\Big.\Big)^{\frac{1}{n-1}}+D_{1}.
\end{equation}
where $D_{1}$ is the constant of the integration.\\
Thus, we write the expression of the slow-roll parameter $\varepsilon(\phi)$ of this model by substituting Eqns.{\color{blue}(22)} and {\color{blue}(24)} into Eqn.{\color{blue}(14)} as
\begin{equation}
\varepsilon(\phi)=\frac{1}{2}\Bigg.\Bigg[\frac{\frac{2}{3}\beta\Big.\Big(\frac{e^{\phi}}{n\beta}\Big.\Big)^{\frac{n}{n-1}}-\frac{1}{3}\phi(\frac{e^{\phi}}{n\beta})^{\frac{1}{n-1}}}{\frac{2\beta(n-1)}{3n}\Big.\Big(\frac{e^{\phi}}{\beta n}\Big.\Big)^{\frac{n}{n-1}}+\frac{(n-1)(n-1+\phi)}{3}\Big.\Big(\frac{e^{\phi}}{\beta n}\Big.\Big)^{\frac{1}{n-1}}+D_{1}}\Bigg.\Bigg]^{2}.
\end{equation}
The expression of the slow-roll $\eta(\phi)$ parameter is obtained by substituting Eqns.{\color{blue}(23)} and {\color{blue}(24)} into Eqn.{\color{blue}(15)} and hence gives
\begin{equation}
\eta(\phi)=\frac{\frac{2\beta n}{3(n-1)}\Big.\Big(\frac{e^{\phi}}{\beta n}\Big.\Big)^{\frac{n}{n-1}}-\frac{\phi}{3(n-1)}\Big.\Big(\frac{e^{\phi}}{\beta}\Big.\Big)^{\frac{1}{n-1}}-\frac{1}{3}\Big.\Big(\frac{e^{\phi}}{\beta n}\Big.\Big)^{\frac{1}{n-1}}}{\frac{2\beta(n-1)}{3n}\Big.\Big(\frac{e^{\phi}}{\beta n}\Big.\Big)^{\frac{n}{n-1}}+\frac{(n-1)(n-1+\phi)}{3}\Big.\Big(\frac{e^{\phi}}{\beta n}\Big.\Big)^{\frac{1}{n-1}}+D_{1}}.
\end{equation}
The spectral index $n_{s}$ and tensor-to-scalar ratio $r$ of  $\beta R^{n}$ model in this study are compiled in Table {\color{blue}1},
with the appropriate choices of $n$, $\beta$ and the scalar field $\phi$.
\begin{table}[ht]
\caption{Table about numerical values of spectral index $n_{s}$ and tensor-to-scalar ratio $r$ for $\beta R^{n}$   using reduced Planck units; for different values of free parameters are listed in this table.} 
\centering 
\begin{tabular}{c c c c c c c c } 
\hline
$n$ & $\beta$ & $D_{1}$ & $\phi$ &  $ r$  & $ n_{s}$ & $r(Planck data)$ \cite{ade2016planck} & $n_{s}(Planck data)$  \cite{ade2016planck} \\ [0.5ex] 

\hline 
0.5 & 1  & 1  & 3 & 0.00210 & 0.96126 & $<$0.11 &0.968 $\pm 0.006$  \\    \hline

0.5 & 1  & 1 & 3.1&  0.00172 & 0.96535 & $<$0.11 &0.968$\pm 0.006$ \\   \hline

0.5 & 1  & 1 & 3.2& 0.00141 & 0.96898 & $<$0.11 &0.968$\pm 0.006$ \\   \hline

0.5 & 1  & 1 & 3.3&  0.00116& 0.97222 & $<$0.11 &0.968$\pm 0.006$  \\    \hline

0.5 &1  &1 & 3.4&  0.00095& 0.97509 & $<$0.11 &0.968$\pm 0.006$ \\   \hline

0.51&2 & 3.5 & 3& 0.00223& 0.95981& $<$0.11 &0.968$\pm 0.006$ \\   \hline

0.51&2 & 3.5 & 3.1&  0.00182& 0.96410& $<$0.11 &0.968$\pm 0.006$ \\   \hline

0.51&2 & 3.5 & 3.2& 0.00148 & 0.96792 & $<$0.11 &0.968$\pm 0.006$ \\   \hline

0.51&2 & 3.5& 3.3& 0.00120& 0.97131 & $<$0.11 &0.968$\pm 0.006$ \\   \hline

0.51&2 & 3.5& 3.4&0.00098 & 0.97433 & $<$0.11 &0.968$\pm 0.006$ \\   \hline

0.52&3 & 7& 3&   0.00244 & 0.95621 & $<$0.11 &0.968$\pm 0.006$ \\   \hline

0.52&3 & 7 & 3.1&0.00197& 0.96106 & $<$0.11 &0.968$\pm 0.006$ \\   \hline

0.52&3 & 7& 3.2& 0.00159 & 0.96535 & $<$0.11 &0.968$\pm 0.006$ \\   \hline

0.52&3 & 7& 3.3& 0.00128& 0.96915  & $<$0.11 &0.968$\pm 0.006$\\   \hline

0.52&3 & 7 & 3.4& 0.00103 &0.97251 & $<$0.11 &0.968$\pm 0.006$ \\ [1ex] 
\hline 
\end{tabular}
\label{table 1} 
\end{table}
\newpage
In this model, we write the $KGE$ {\color{blue}(12)} as
\begin{equation}
\ddot{\phi}+\frac{2\beta}{3}\Big.\Big(  \frac{1}{n\beta} \Big.\Big)^{\frac{n}{n-1}}e^{\frac{n\phi}{n-1}}-\frac{1}{3}\Big.\Big(\frac{1}{n\beta}\Big.\Big)^{\frac{1}{n-1}}e^{\frac{n\phi}{n-1}}=0.
\end{equation}
The numerical solution to Eqn.{\color{blue}(27)} is presented in Figure {\color{blue}1}.  Note that the initial values used while plotting are taken from Table {\color{blue}(1)}.
\begin{figure}[H]
  \centering
\includegraphics[height=8cm,width=13cm]{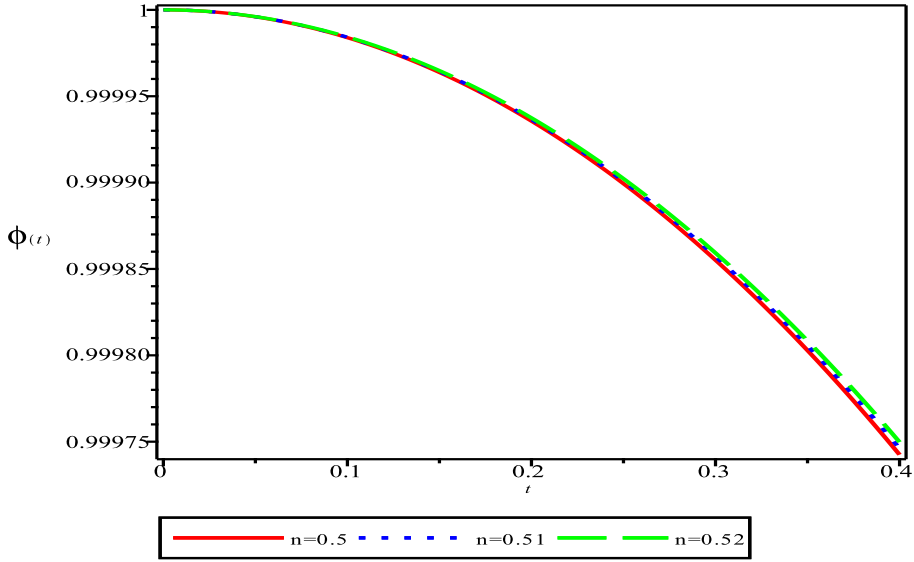}
\caption{The plot shows the numerical solution of Eqn. (27)  using reduced Planck units of Klein-Gordon equation for $\beta R^{n}$ model. We used $\beta=1$, $n=0.5$, $n=0.51$, $n=0.52$, $\phi(0)=3$ and $\phi'(0)=0$. Note that there is a decrease of scalar field with the increase of time.}
\label{fig:1}
\end{figure}
\subsection{Model2: $f(R)=\alpha R+\beta R^{n}$}
Among various $f(R)$ models,  we now  consider the case of polynomial $f(R)$ model as described in \cite{barrow1983stability} as
\begin{equation}
f(R)=\alpha R+\beta R^{n}.
\end{equation}
This model has been one of the most popular fourth-order gravitational theories with power-law corrections. It is a generalisation of both to $GR$ and the $R^{n}$ actions since $\alpha=1$, $\beta=0$ reduces to $GR$ and $\alpha=0$ reduces to $R^{n}$ case. In our study, the parameter $\alpha$, $\beta$ and $n$ will be taken to be real. This is the first model of inflation proposed by Starobinsky in 1980 \cite{starobinsky1980new}, it was also found that, this model is well consistent with the temperature anisotropies observed in $CMB$ and thus it can be a viable alternative to the scalar-field models of inflation. This toy  model is currently gaining popularity as an  alternative model of gravitation within the context of early and late universe \cite{carloni2006bounce}.
Using the definition of the scalar field, this mode has the following form
\begin{equation}
f(\phi)=\alpha\Big.\Big(\frac{e^{\phi}-\alpha}{n\beta}\Big.\Big)^{\frac{1}{n-1}}+\beta\Big.\Big(\frac{e^{\phi}-\alpha}{\beta n}\Big.\Big)^{\frac{n}{n-1}}.
\end{equation}
Thus, we write the effective scalar field potential Eqn.{\color{blue}(13)} to this model as
\begin{equation}
V'(\phi)=\frac{2\alpha}{3}\Bigg.\Bigg(\frac{e^{\phi}-\alpha}{\beta n}\Bigg.\Bigg)^{\frac{1}{n-1}}+\frac{2\beta}{3}\Bigg.\Bigg(\frac{e^{\phi}-\alpha}{\beta n}\Bigg.\Bigg)^{\frac{n}{n-1}}-\phi \Bigg.\Bigg(\frac{e^{\phi}-\alpha}{\beta n}\Bigg.\Bigg)^{\frac{1}{n-1}}.
\end{equation}
The second derivative of Eqn.{\color{blue}(30)} gives
\begin{align}
\begin{split}
V''(\phi)&=\frac{2n\beta e^{\phi}}{3(n-1)(e^{\phi}-\alpha)}\Bigg.\Bigg(\frac{e^{\phi}-\alpha}{\beta n}\Bigg.\Bigg)^{\frac{n}{n-1}}+\frac{2\alpha e^{\phi}}{3(n-1)(e^{\phi}-\alpha)}\Bigg.\Bigg(\frac{e^{\phi}-\alpha}{\beta n}\Bigg.\Bigg)^{\frac{1}{n-1}}-\frac{1}{3}\Bigg.\Bigg(\frac{e^{\phi}-\alpha}{\beta n}\Bigg.\Bigg)^{\frac{1}{n-1}}\\&
-\frac{\phi e^{\phi}}{3(n-1)(e^{\phi}-\alpha)}\Bigg.\Bigg(\frac{e^{\phi}-\alpha}{\beta n}\Bigg.\Bigg)^{\frac{1}{n-1}}.
\end{split}
\end{align}
Thus , the potential has the form
\begin{equation}
V(\phi)=\frac{2\alpha}{3} \int \Bigg.\Bigg(\frac{e^{\phi}-\alpha}{n\beta} \Bigg.\Bigg)^{\frac{1}{n-1}}d\phi+\frac{2\beta}{3} \int  \Bigg.\Bigg(\frac{e^{\phi}-\alpha}{n\beta} \Bigg.\Bigg)^{\frac{n}{n-1}}d\phi-\frac{\phi}{3} \int \Bigg.\Bigg(\frac{e^{\phi}-\alpha}{n\beta} \Bigg.\Bigg)^{\frac{1}{n-1}}d\phi.
\end{equation}
The following couple of equations are expressions of the slow-roll parameters $\eta(\phi)$ and $\varepsilon{\phi}$ respectively
\begin{multline}
\eta(\phi)=\frac{\frac{2n\beta e^{\phi}}{3(n-1)(e^{\phi}-\alpha)}\Bigg.\Bigg(\frac{e^{\phi}-\alpha}{\beta n}\Bigg.\Bigg)^{\frac{n}{n-1}}+\frac{2\alpha e^{\phi}}{3(n-1)(e^{\phi}-\alpha)}\Bigg.\Bigg(\frac{e^{\phi}-\alpha}{\beta n}\Bigg.\Bigg)^{\frac{1}{n-1}}-\frac{1}{3}\Bigg.\Bigg(\frac{e^{\phi}-\alpha}{\beta n}\Bigg.\Bigg)^{\frac{1}{n-1}}}{\frac{2\alpha}{3} \int \Bigg.\Bigg(\frac{e^{\phi}-\alpha}{n\beta} \Bigg.\Bigg)^{\frac{1}{n-1}}d\phi+\frac{2\beta}{3} \int  \Bigg.\Bigg(\frac{e^{\phi}-\alpha}{n\beta} \Bigg.\Bigg)^{\frac{n}{n-1}}d\phi-\frac{\phi}{3} \int \Bigg.\Bigg(\frac{e^{\phi}-\alpha}{n\beta} \Bigg.\Bigg)^{\frac{1}{n-1}}d\phi}\\-\frac{\frac{\phi e^{\phi}}{3(n-1)(e^{\phi}-\alpha)}\Bigg.\Bigg(\frac{e^{\phi}-\alpha}{\beta n}\Bigg.\Bigg)^{\frac{1}{n-1}}
}{\frac{2\alpha}{3} \int \Bigg.\Bigg(\frac{e^{\phi}-\alpha}{n\beta} \Bigg.\Bigg)^{\frac{1}{n-1}}d\phi+\frac{2\beta}{3} \int  \Bigg.\Bigg(\frac{e^{\phi}-\alpha}{n\beta} \Bigg.\Bigg)^{\frac{n}{n-1}}d\phi-\frac{\phi}{3} \int \Bigg.\Bigg(\frac{e^{\phi}-\alpha}{n\beta} \Bigg.\Bigg)^{\frac{1}{n-1}}d\phi}
\end{multline}
\begin{equation}
\varepsilon(\phi)=\frac{1}{2}\frac{\Bigg.\Bigg[\frac{2\alpha}{3}\Bigg.\Bigg(\frac{e^{\phi}-\alpha}{\beta n}\Bigg.\Bigg)^{\frac{1}{n-1}}+\frac{2\beta}{3}\Bigg.\Bigg(\frac{e^{\phi}-\alpha}{\beta n}\Bigg.\Bigg)^{\frac{n}{n-1}}-\phi \Bigg.\Bigg(\frac{e^{\phi}-\alpha}{\beta n}\Bigg.\Bigg)^{\frac{1}{n-1}} \Bigg.\Bigg]^{2}}{\Bigg.\Bigg[[\frac{2\alpha}{3} \int \Bigg.\Bigg(\frac{e^{\phi}-\alpha}{n\beta} \Bigg.\Bigg)^{\frac{1}{n-1}}d\phi+\frac{2\beta}{3} \int  \Bigg.\Bigg(\frac{e^{\phi}-\alpha}{n\beta} \Bigg.\Bigg)^{\frac{n}{n-1}}d\phi-\frac{\phi}{3} \int \Bigg.\Bigg(\frac{e^{\phi}-\alpha}{n\beta} \Bigg.\Bigg)^{\frac{1}{n-1}}d\phi\Bigg.\Bigg]^{2}}.
\end{equation}
\begin{table}[ht]
\caption{Table about numerical values of spectral index $n_{s}$ and tensor-to-scalar ratio $r$  for $f(R)=\alpha R+\beta R^{n}$   using reduced Planck units.} 
\centering 
\begin{tabular}{c c c c c c c c } 
\hline
$n$ & $\beta$ & $\alpha$ &$\phi$ &  $ r$  & $ n_{s}$ & $r(Planck data)$  \cite{ade2016planck}& $n_{s}(Planck data)$ \cite{ade2016planck} \\ [0.5ex] 

\hline 
0.5 & 0.002 & 0.1& 1.4 & 0.00237 & 0.96271 & $<$0.11 &0.968$\pm 0.006$  \\   \hline

0.5 & 0.002 & 0.1& 1.5 & 0.00193 & 0.96672 & $<$0.11 &0.968$\pm 0.006$  \\   \hline

0.5 & 0.002 & 0.1& 1.6 & 0.00158&0.97026 & $<$0.11 &0.968$\pm 0.006$ \\   \hline

0.5 & 0.002 & 0.1&  1.7 & 0.00129 & 0.97339& $<$0.11 &0.968$\pm 0.006$  \\   \hline

0.5 & 0.002& 0.1& 1.8 &0.00105 & 0.97617 & $<$0.11 &0.968$\pm 0.006$ \\   \hline

0.5 & 0.005 & 0.001& 2.6 &0.00164 & 0.96913 & $<$0.11 &0.968$\pm 0.006$  \\    \hline

0.5 & 0.005 & 0.001& 2.65 & 0.00148 & 0.97079 & $<$0.11 &0.968$\pm 0.006$ \\   \hline

0.5 & 0.005 & 0.001& 2.7 & 0.00133 &0.97235 & $<$0.11 &0.968$\pm 0.006$ \\   \hline

0.5 & 0.005 & 0.001&  2.75 & 0.00120 &0.97383 & $<$0.11 &0.968$\pm 0.006$ \\   \hline

0.5 & 0.005 & 0.dr001& 2.8 & 0.00108 & 0.97522 & $<$0.11 &0.968$\pm 0.006$ \\   [1ex]

\hline 
\end{tabular}
\label{table 2} 
\end{table}
\newpage
The $KGE$ {\color{blue}(12)} of this model is therefore has the form
\begin{equation}
\ddot{\phi}+\frac{1}{3}\Bigg.\Bigg[\frac{2\alpha}{(\beta n)^{\frac{1}{n-1}}}\Big.\Big(e^{\phi}-\alpha \Big.\Big)^{\frac{1}{n-1}}+\frac{2\beta}{(n\beta)^{\frac{n}{n-1}}}\Big.\Big(e^{\phi}-\alpha\Big.\Big)^{\frac{n}{n-1}}-\frac{1}{(n\beta)^{\frac{1}{n-1}}}e^{\phi}\Big.\Big(e^{\phi}-\alpha\Big.\Big)^{\frac{1}{n-1}}\Bigg.\Bigg]=0.
\end{equation}
The numerical solution to Eqn.{\color{blue}(35)} is presented in Figure {\color{blue}2}. Note that the initial values used while plotting are taken from Table {\color{blue}(2)}.
\begin{figure}[H]
  \centering
\includegraphics[height=8cm,width=13cm]{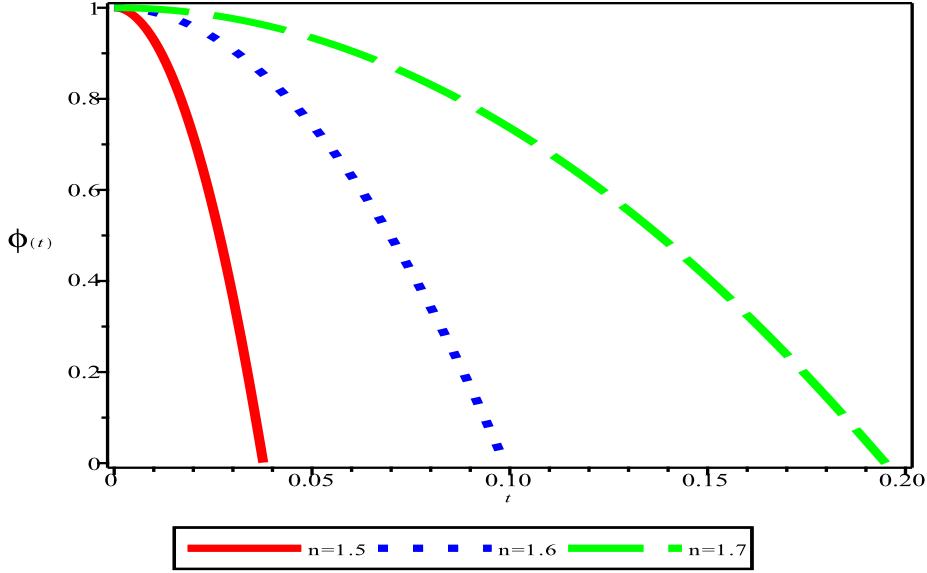}
\caption{The plot shows the evolution of scalar field obtained by solving numerically Eqn.{\color{blue}(35)} for $n=1.5$, $n=1.6$ and $n=1.7$, $\beta=0.002$ and $\alpha=0.001$. We solved the Klein-Gordon equation setting the initial conditions to be
$\dot{\phi}(t_{0})=1.7$ and $\phi(t_{0})=0$   and reduced Planck units were used. }
\label{fig:2}
\end{figure}
\subsection{Model3: $f(R)=R+\frac{\mu^{2n+2}}{(-R)^{n}}$}
This model is similar to the previous except that here we have the inverse power-law as additional term to the usual $GR$ concept and it is defined as
\begin{equation}
f(R)=R+\frac{\mu^{2n+2}}{(-R)^{n}}.
\end{equation}
The model we are considering now is driven by a constant parameter $\mu$ which is normalised to unit. It means that if $\mu = 0$, we  immediately turn to our $ GR$; see in  \cite{li2007cosmology}. This model in terms of scalar field has the the following form
\begin{equation}
f(\phi)=-\Big.\Big(\frac{n\mu^{2n+2}}{e^{\phi-1}}\Big.\Big)^{\frac{1}{n+1}}+\mu^{2n+2}(\frac{e^{\phi}-1}{n\mu^{2n+2}})^{\frac{n}{n+1}}.
\end{equation}
The derivative of the potential, $V'(\phi)$ obtained from Eqn.{\color{blue}(13)} is
\begin{equation}
V'(\phi)=\frac{1}{3}\Big.\Big[-2\Big.\Big(\frac{n\mu^{2n+2}}{e^{\phi}-1}\Big.\Big)^{\frac{1}{n+1}}+\mu^{2n+2}\Big.\Big(\frac{e^{\phi}-1}{n\mu^{2n+2}}\Big.\Big)^{\frac{n}{n+1}}]+\phi\Big.\Big(\frac{n\mu^{2n+2}}{e^{\phi}-1}\Big.\Big)^{\frac{1}{n+1}}.
\end{equation}
The second derivative of Eqn.{\color{blue}(38)} has the following form
 \begin{multline}
 V''(\phi) = \frac{1}{3}[-K_{1}(n+2)(\phi-2)(1-e^{\phi})^{-(n+3)} + K_{1}(1-e^{\phi})^{-(n+2)} \\+(n+2)(n+1)K_{2}(1-e^{\phi})^{(n+2)(n+1) -1}],
 \end{multline}
 where
\begin{equation}
K_{1} = \mu^{2n+2}(n+1)^{n+2},
\end{equation}
and
\begin{equation}
K_{2} = \frac{1}{[(n+1)\mu^{2n+2}]^{(n+2)(n)+1)}}.
\end{equation}
Using the same definition as Eqn.{\color{blue}(14,15)} we have the expression of the slow-roll parameters $\varepsilon{\phi}$ and $ \eta(\phi) $ as
  \begin{equation}
  \varepsilon(\phi)  = \frac{1}{6}\frac{\Big.\Big[K_{1}(1-e^{\phi})^{-(n+2)}(\phi -2) +K_{2} (1-e^{\phi})^{(n+2)(n+1)}\Big.\Big]^{2}} {\Big.\Big[K_{1}\int(1-e^{\phi})^{-(n+2)}(\phi -2)d\phi + K_{2}\int(1-e^{\phi})^{-(n+2)}d\phi\Big.\Big]^{2}},
  \end{equation}
  \begin{multline}
  \eta(\phi) =\frac{1}{3\kappa}(\frac{V^{''}}{V})
  = \frac{\frac{1}{k}\Big.\Big[-K_{1}(n+2)(\phi-2)(1-e^{\phi})^{-(n+3)} \Big.\Big]}{\Big.\Big[K_{1}\int(1-e^{\phi})^{-(n+2)}(\phi -2)d\phi + K_{2}\int(1-e^{\phi})^{-(n+2)}d\phi\Big.\Big]^{2}}\\+\frac{ \frac{1}{k}\Big.\Big[K_{1}(1-e^{\phi})^{-(n+2)} +(n+2)(n+1)K_{2}(1-e^{\phi})^{[(n+2)(n+1) -1]}\Big.\Big]}{\Big.\Big[K_{1}\int(1-e^{\phi})^{-(n+2)}(\phi -2)d\phi + K_{2}\int(1-e^{\phi})^{-(n+2)}d\phi\Big.\Big]^{2}}
  \end{multline}
\begin{table}[ht]
\caption{Table about numerical values of spectral index $n_{s}$ and tensor-to-scalar ratio $r$ for $f(R)=R+\frac{\mu^{2n+2}}{(-R)^{n}}$   using reduced Planck units. } 
\centering 
\begin{tabular}{c c c c c c c} 
\hline
$n$ & $\mu$ & $\phi$ &  $ r $ & $ n_{s}$ & $r(Planck data)$  \cite{ade2016planck} & $n_{s}(Planck data)$  \cite{ade2016planck} \\ [0.5ex] 

\hline 
1.066 & 1 & 0.555& 0.01782 & 0.97347& $<$0.11 &0.968$\pm 0.006$ \\    \hline

1.066 & 1 & 0.556& 0.01825 & 0.97223 & $<$0.11 &0.968$\pm 0.006$ \\   \hline

1.066 & 1 & 0.557& 0.01870 & 0.97098 & $<$0.11 &0.968$\pm 0.006$  \\   \hline

1.066 & 1 & 0.558& 0.01914&0.96974 & $<$0.11 &0.968$\pm 0.006$  \\    \hline

1.066 &  1 & 0.559& 0.01959 & 0.96849 & $<$0.11 &0.968$\pm 0.006$  \\  \hline

1.059& 1.04& 0.678 & 0.15746 & 0.97341 & $<$0.11 &0.968$\pm 0.006$ \\   \hline

1.059 & 1.04& 0.679 & 0.15891 &0.97169 & $<$0.11 &0.968$\pm 0.006$ \\   \hline

1.059 &1.04 & 0.680 & 0.16037 & 0.96996 & $<$0.11 &0.968$\pm 0.006$ \\   \hline

1.059 & 1.04 & 0.681 & 0.16183 & 0.96822 & $<$0.11 &0.968$\pm 0.006$ \\   \hline

1.059 & 1.04 & 0.682 & 0.16330 &0.96648 & $<$0.11 &0.968$\pm 0.006$  \\   \hline

1.051 &1.01 & 0.563 & 0.02996 & 0.97128 & $<$0.11 &0.968$\pm 0.006$  \\   \hline

1.051 & 1.01 & 0.564 & 0.03057 &  0.97000 & $<$0.11 &0.968$\pm 0.006$  \\  \hline

1.051 & 1.01& 0.565 & 0.03118 & 0.96871 & $<$0.11 &0.968$\pm 0.006$ \\   \hline

1.051 & 1.01&0.566 & 0.03180 & 0.96742& $<$0.11 &0.968$\pm 0.006$ \\   \hline

1.068 & 1.01&0.567& 0.03242 & 0.96613 & $<$0.11 &0.968$\pm 0.006$ \\ [1ex] 
\hline 
\end{tabular}
\label{table 3} 
\end{table}
In this model, we write the $KGE$ {\color{blue}(12)} as
\begin{equation}
\ddot{\phi}+\frac{1}{3}\Bigg.\Bigg(A_{3}\Big.\Big(e^{\phi}-1\Big.\Big)^{\frac{n}{n+1}}+\Big.\Big(Ce^{\phi}-B_{3}\Big.\Big)\Big.\Big(e^{\phi}-1\Big.\Big)^{\frac{-1}{n+1}}\Bigg.\Bigg)=0,
\end{equation}
where
\begin{equation}
A_{3}=2\mu^{2n+2} \Big.\Big(n\mu^{2n+2}\Big.\Big)^{\frac{-n}{n+1}},
\end{equation}
\begin{equation}
B_{3}=2\Big.\Big(n\mu^{2n+2}\Big.\Big)^{\frac{1}{n+1}},
\end{equation}
\begin{equation}
C=\Big.\Big(n\mu^{2n+2}\Big.\Big)^{\frac{1}{n+1}}.
\end{equation}
\newpage
The numerical solution to Eqn.{\color{blue}(44)} is presented in Figure {\color{blue}3}.  Note that the initial values used while plotting are taken from Table {\color{blue}(3)}.
\begin{figure}[H]
  \centering
\includegraphics[height=8cm,width=13cm]{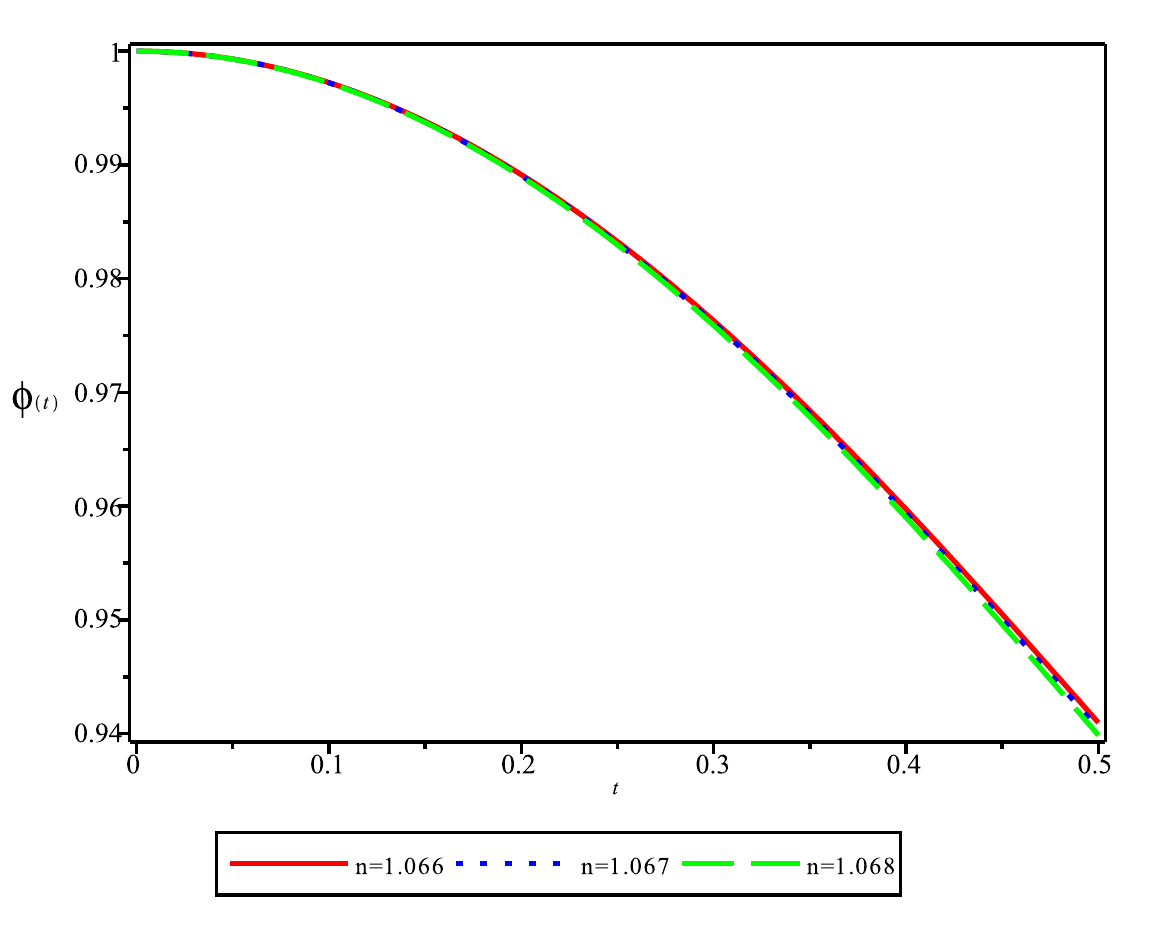}
\caption{The evolution of scalar field as function of time of $(R)=R+\frac{\mu^{2n+2}}{(-R)^{n}}$ mode is plotted in this figure. It is a result of numerical solution of Eqn.{\color{blue}(44)}.
The free parameters used are $\mu=1$, $n=1.066$, $n=1.067$, $n=1.068$, $\phi(t=0)=0.553$ and $\phi'(t=0)=0$   and reduced Planck units were used.}
\label{fig:3}
\end{figure}
\subsection{Model4: $f(R)=\alpha e^{\lambda R}$}
We consider theories with a Lagrangians which may be expressed as an exponential of the Ricci scalar as 
\begin{equation}
f(R)=\alpha e^{\lambda R}.
\end{equation}
This model was widely treated because it is easier and does not present complications for any one who wants to analyse  the stability of cosmological inflation using the dynamical system  analysis \cite{carloni2006bounce}. This type of Lagrangian is interesting because it contains both $R^{n}$ and $\alpha R+\beta R^{n}$ models due to the fact that the exponential can be developed in powers of the Ricci scalar. Using Taylor expansion, the polynomial $f(R)$ can be obtained and hence, its reduction to $GR$ can be achieved. in our analysis $\alpha$ and $\lambda$ were taken to be positive values. Based on the definition of the scalar field, this model has the form
\begin{equation}
f(\phi)=\alpha e^{\phi-ln(\alpha \lambda)}.
\end{equation}
An expression  for the derivative of the potential, $V'(\phi)$  has the form
\begin{equation}
V'(\phi)=\frac{2\alpha}{3}e^{\phi-ln(\alpha \lambda)}-\frac{1}{3\lambda}e^{\phi}\Big.\Big(\phi-ln(\alpha\lambda)\Big.\Big).
\end{equation}
The second derivative of Eqn.{\color{blue}(50)} gives
\begin{equation}
V''(\phi)=\frac{2\alpha e^{\phi-ln(\alpha \lambda)}}{3}-\frac{e^{\phi}\Big.\Big(\phi-ln(\alpha \lambda)\Big.\Big)}{3\lambda}-\frac{e^{\phi}}{3\lambda}.
\end{equation}
Thus, the potential is given by integration of the Eqn. {\color{blue} (50)}
\begin{equation}
V(\phi)=\frac{2\alpha e^{\phi-ln(\alpha \lambda)}}{3}+\frac{e^{\phi}\Big.\Big(-\phi+ln(\alpha\lambda)+1\Big.\Big)}{3\lambda}+D_{2}.
\end{equation}
Using the same definition given by Eqn.{\color{blue} (14)}, we have the expression of $\varepsilon(\phi)$ as
\begin{equation}
\varepsilon(\phi)=\frac{\frac{2\alpha e^{\phi-ln(\alpha \lambda)}}{3}-\frac{e^{\phi}\Big.\Big(\phi-ln(\alpha \lambda)\Big.\Big)}{3\lambda}-\frac{e^{\phi}}{3\lambda}}{\frac{2\alpha e^{\phi-ln(\alpha \lambda)}}{3}+\frac{e^{\phi}\Big.\Big(-\phi+ln(\alpha\lambda)+1\Big.\Big)}{3\lambda}+D_{2}}.
\end{equation}
Using the same definition of the Eqn.{\color{blue}(15)} , we have the expression of $\eta(\phi)$ as
\begin{equation}
\eta(\phi)=\frac{\Bigg.\Bigg[\frac{2\alpha}{3}e^{\phi-ln(\alpha \lambda)}-\frac{1}{3\lambda}e^{\phi}\Big.\Big(\phi-ln(\alpha\lambda)\Big.\Big)\Bigg.\Bigg]^{\frac{1}{2}}}{\Bigg.\Bigg[\frac{2\alpha e^{\phi-ln(\alpha \lambda)}}{3}+\frac{e^{\phi}\Big.\Big(-\phi+ln(\alpha\lambda)+1\Big.\Big)}{3\lambda}+D_{2}\Bigg.\Bigg]^{\frac{1}{2}}}.
\end{equation}
\begin{table}[ht]
\caption{Table about numerical values of spectral index $n_{s}$ and tensor-to-scalar ratio $r$ for $f(R)=\alpha e^{\lambda R}$   using reduced Planck units.} 
\centering 
\begin{tabular}{c c c c c c c c} 
\hline
  $ \alpha$  &  $\lambda$ & $D_{2}$& $\phi$ & $ r $ & $ n_{s}$  \cite{ade2016planck}& $r(Planck data)$ & $n_{s}(Planck data)$ \cite{ade2016planck} \\ [0.5ex] 

\hline 
 0.002 & 1.1& 40& 1.1 & 0.00189  & 0.96129 & $<$0.11 &0.968$\pm 0.006$   \\    \hline

 0.002 & 1.1 & 50 & 1.2 & 0.00091  & 0.96600 & $<$0.11 &0.968$\pm 0.006$ \\   \hline

 0.002 & 1.1 & 55  & 1.3 & 0.00077 & 0.96960 & $<$0.11 &0.968$\pm 0.006$  \\   \hline

 0.002 & 1.1 & 35 & 1.4 &0.00480 & 0.96555 & $<$0.11 &0.968$\pm 0.006$  \\    \hline

 0.002 & 1.1& 30  & 1.5 & 0.00626 & 0.97096 & $<$0.11 &0.968$\pm 0.006$ \\    \hline

 0.002 & 1.01 & 40 & 1.1 & 0.00250 & 0.95568 & $<$0.11 &0.968$\pm 0.006$  \\   \hline

 0.002 & 1.01 & 50  & 1.2 & 0.00120 & 0.96147 & $<$0.11 &0.968$\pm 0.006$ \\   \hline

 0.002 & 1.01 & 55  & 1.3 & 0.00101 & 0.96555 & $<$0.11 &0.968$\pm 0.006$ \\   \hline

 0.002 & 1.01 & 35  & 1.4 & 0.00513 & 0.96415 & $<$0.11 &0.968$\pm 0.006$ \\   \hline

 0.002 & 1.01 & 30  & 1.5 &  0.00845 & 0.96862 &$ <$0.11 &0.968$\pm 0.006$ \\   \hline

 0.002 & 1.02 & 40  & 1.1 & 0.00242 & 0.95637 & $<$0.11 &0.968$\pm 0.006$ \\   \hline

 0.002 & 1.02 & 50  & 1.2 & 0.00116 & 0.96202 & $<$0.11 &0.968$\pm 0.006$ \\   \hline

 0.002 & 1.02 & 55 &1.3 & 0.00098  & 0.96604  & $<$0.11 &0.968$\pm 0.006$\\   \hline

 0.002 & 1.02 & 35  & 1.4 & 0.00496 & 0.96486 & $<$0.11 &0.968$\pm 0.006$ \\   \hline

 0.002 & 1.02 & 30  & 1.5 &0.00816  & 0.96943 & $<$0.11 &0.968$\pm 0.006$\\[1ex]
\hline 
\end{tabular}
\label{table 4} 
\end{table}
The $KGE$ {\color{blue}(12)} in this model, has the following the form
\begin{equation}
\ddot{\phi}+\frac{2\alpha e^{-\ln(\alpha\lambda)}-\frac{1}{\lambda}\ln(\alpha\lambda)e^{\phi}}{3}-\frac{\phi e^{\phi}}{3\lambda}=0,
\end{equation}
\newpage
The numerical solution to Eqn.{\color{blue}(55)} is presented in Figure {\color{blue}4}.
\begin{figure}[H]
  \centering
\includegraphics[height=8cm,width=13cm]{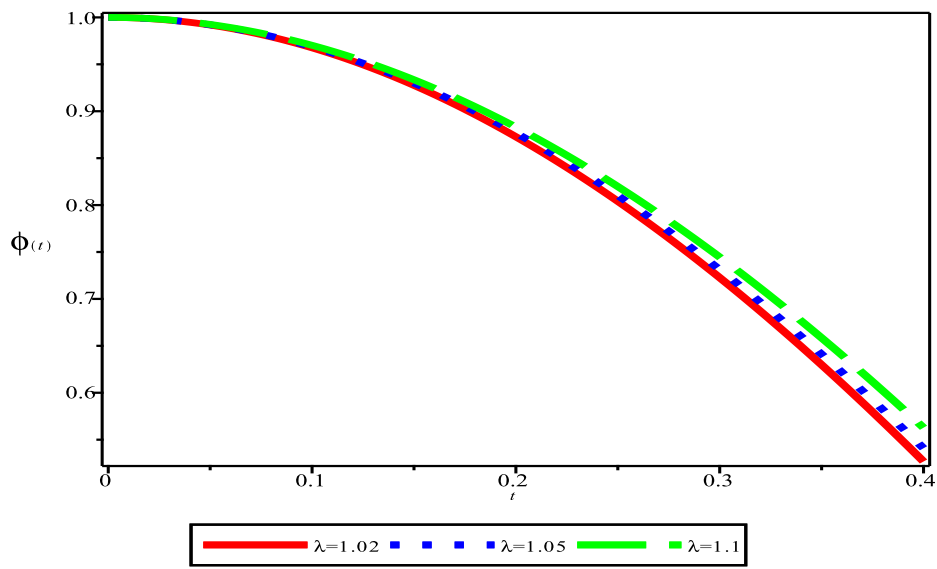}
\caption{The numerical solution of Eqn.{\color{blue}(55)} is presented in this figure. The free parameters used are $\alpha=0.002$, $\lambda=1.02$, $\lambda=1.05$ and $\lambda=1.1$. Setting the initial conditions to be $\phi(t=0)=1.5$ and $\phi'(t=0)=0$ and    reduced Planck units were used.}
\label{fig:4}
\end{figure}
\subsection{Model5: $R-\frac{\mu^{4}}{R}$}
This model is a special case of model 3. Only we are taking $n =1$ and thus the approach and way of treatment are quite the same. 
\begin{equation}
f(R)=R-\frac{\mu^{4}}{R}.
\end{equation}
It is interesting to note that when $\mu=0$, this model reduces to $GR$ case  \cite{faulkner2007constraining}.
In this model, the function $f(R)$ as function of scalar field  $\phi$ is given as
\begin{equation}
f(\phi)=\mu^{2}\Big.\Big[\Big.\Big(e^{\phi}-1\Big.\Big)^{\frac{-1}{2}}-\Big.\Big(e^{\phi}-1\Big.\Big)^{\frac{1}{2}}\Big.\Big].
\end{equation}
The expression of the derivative of the effective potential of this model has the form
\begin{equation}
V'(\phi)=\frac{1}{3}\Big.\Big[2\mu^{2}\Big.\Big(e^{\phi}-1\Big.\Big)^{\frac{-1}{2}}-\Big.\Big(e^{\phi}-1\Big.\Big)^{\frac{1}{2}}-\mu^{2}\phi\Big.\Big(e^{\phi}-1\Big.\Big)\Big.\Big].
\end{equation}
The second derivative of Eqn.{\color{blue}(58)} yields
\begin{equation}
V''(\phi)=\frac{1}{3}\Big.\Big[-\mu^{2}e^{\phi}(e^{\phi}-1)^{-\frac{3}{2}}-\frac{e^{\phi}}{2}(e^{\phi}-1)^{\frac{-1}{2}}-\mu^{2}(e^{\phi}-1)+\mu^{2}\phi e^{\phi}\Big.\Big].
\end{equation}
By integration the Eqn.{\color{blue}(58)} with respect to $\phi$, we have the potential $V'(\phi)$
\begin{equation}
V(\phi)=\frac{2\mu^{2}}{3}arctan\sqrt{e^{\phi}-1}-\frac{4\mu^{2}}{3}\sqrt{e^{\phi}-1}+\frac{4}{3}\mu^{2}arctan\sqrt{e^{\phi}-1}-\frac{2\mu^{2}}{3}arctan\sqrt{e^{\phi}-1}+D_{3}.
\end{equation}
The potential slow-roll parameters $\varepsilon(\phi) $ and $\eta(\phi)$ for this model are given by
\begin{equation}
\varepsilon(\phi)=\frac{\frac{1}{3}\Bigg.\Bigg[\frac{-\mu^{2}e^{\phi}}{(e^{\phi}-1)^{\frac{3}{2}}}-\frac{-\mu^{2}e^{\phi}}{(e^{\phi}-1)^{\frac{1}{2}}}-\frac{-\mu^{2}}{(e^{\phi}-1)^{\frac{1}{2}}}-\frac{-\mu^{2}e^{\phi}\phi}{(e^{\phi}-1)^{\frac{3}{2}}}\Bigg.\Bigg]}{\frac{2\mu^{2}}{3}\arctan{\sqrt{e^{\phi}-1}}-\frac{4\mu^{2}}{3}\sqrt{e^{\phi}-1}+\frac{4}{3}\mu^{2}\arctan{\sqrt{e^{\phi}-1}}-\frac{2\mu^{2}}{3}\arctan{\sqrt{e^{\phi}-1}}+D_{3}},
\end{equation}
\begin{equation}
\eta(\phi)=\frac{1}{2}\Bigg.\Bigg[\frac{\frac{2\mu^{2}}{3(e^{\phi}-1)^{\frac{1}{2}}}-\frac{2\mu^{2}}{3}(e^{\phi}-1)^{\frac{1}{2}}-\frac{\phi \mu^{2}}{3(e^{\phi}-1)^{\frac{1}{2}}}}{\frac{2\mu^{2}}{3}\arctan{\sqrt{e^{\phi}-1}}-\frac{4\mu^{2}}{3}\sqrt{e^{\phi}-1}+\frac{4}{3}\mu^{2}\arctan{\sqrt{e^{\phi}-1}}-\frac{2\mu^{2}}{3}\arctan{\sqrt{e^{\phi}-1}}+D_{3}}\Bigg.\Bigg]^{2}.
\end{equation}
\begin{table}[ht]
\caption{Table about numerical values of spectral index $n_{s}$ and tensor-to-scalar ratio $r$ for $f(R)=R-\frac{\mu^{4}}{R}$   using reduced Planck units.} 
\centering 
\begin{tabular}{c c c c c c c } 
\hline
$\mu$  & $D_{3}$ & $\phi$& $ r $ & $ n_{s}$ & $r(Planck data)$  \cite{ade2016planck}& $n_{s} (Planck data)$ \cite{ade2016planck} \\ [0.5ex] 

\hline 
0.7 &  18 & 1.1 & 0.00888 & 0.96184 & $<$0.11 &0.968$\pm 0.006$ \\   \hline

0.7 & 18 & 1.2 & 0.01381 &0.96760 & $<$0.11 &0.968$\pm 0.006$ \\   \hline

0.7 &  18 & 1.25  & 0.01694 & 0.97052 & $<$0.11 &0.968$\pm 0.006$ \\   \hline

0.7 & 18 & 1.3 & 0.02062 & 0.97347 & $<$0.11 &0.968$\pm 0.006$ \\   \hline

0.8 & 19 & 1.1 & 0.00649 & 0.96600& $<$0.11 &0.968$\pm 0.006$ \\   \hline

0.8  & 19 & 1.2 & 0.01478 & 0.97261 & $<$0.11 &0.968$\pm 0.006$ \\   \hline

0.8  & 19 & 1.23 & 0.0169 & 0.97488 & $<$0.11 &0.968$\pm 0.006$ \\   \hline

0.8  & 19 & 1.24  & 0.01776 & 0.97565 & $<$0.11 &0.968$\pm 0.006$ \\   \hline

0.5 &  20 & 1.1 & 0.00586 &0.96423 & $<$0.11 &0.968$\pm 0.006$ \\    \hline

0.5  & 20 & 1.2 & 0.00808 &0.96641 & $<$0.11 &0.968$\pm 0.006$ \\   \hline

0.5  & 20 & 1.3 & 0.01095&0.96864 & $<$0.11 &0.968$\pm 0.006$  \\   \hline

0.5  & 20 & 1.4 &0.01465 &  0.97102 & $<$0.11 &0.968$\pm 0.006$ \\  [1ex] 
\hline 
\end{tabular}
\label{table 5} 
\end{table}
\newpage
In this model, we write the $KGE$ {\color{blue}(12)} as
\begin{equation}
\ddot{\phi}-\frac{1}{3}\Bigg.\Bigg[2\Big.\Big(\frac{\mu^{4}}{e^{\phi}-1}\Big.\Big)^{\frac{1}{2}}-2\mu^{4}\Big.\Big(\frac{e^{\phi}-1}{\mu^{4}}\Big.\Big)^{\frac{1}{2}}-e^{\phi}\Big.\Big(\frac{\mu^{4}}{e^{\phi}-1}\Big.\Big)^{\frac{1}{2}}\Bigg.\Bigg]=0.
\end{equation}
The numerical solution to Eqn.{\color{blue}(63)} is presented in Figure {\color{blue}5}. The behaviour  of solution  is decreasing with the increase of time.
\begin{figure}[H]
  \centering
\includegraphics[height=8cm,width=13cm]{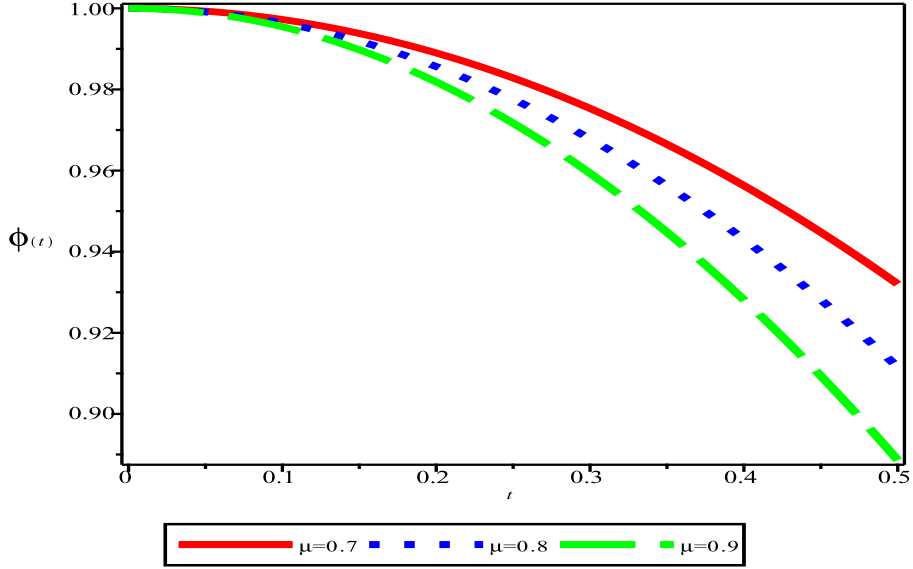}
\caption{This plot shows the evolution of scalar field. It is obtained by solving numerically Eqn.{\color{blue}(63)}; for $\mu=0.7$, $\mu=0.8$, $\mu=0.9$, $\phi(t=0)=1.5$ and $\phi'(t=0)=0$  and reduced Planck units were used.}
\label{fig:5}
\end{figure}
\subsection{Model6: $f(R)=R-(1-n) \mu^{2}\frac{R^{n}}{\mu^{2n}}$}
We consider the case where $f(R)$ model is given by 
\begin{equation}
f(R)=R-(1-n) \mu^{2}\frac{R^{n}}{\mu^{2n}}.
\end{equation}
This is a simple, polynomial $f(R)$ function that quickly reduces to the Einstein General theory of Relativity  for $n=1$ see in  \cite{carroll2004cosmic,faulkner2007constraining}. it successfully achieves late-time acceleration as the $\frac{\mu}{R}$ term starts to dominate.
This $f(R)$ model has the following form in terms of scalar field 
\begin{equation}
f(\phi)=\Bigg.\Bigg(\frac{\mu^{2n-2}(e^{\phi}-1)}{n-n^{2}}\Bigg.\Bigg)^{\frac{1}{n-1}}- \frac{(1-n)\mu^{2}}{\mu^{2n}}\Bigg.\Bigg( \frac{\mu^{2n-2}(e^{\phi}-1)}{n-n^{2}}  \Bigg.\Bigg)^{\frac{n}{n-1}}.
\end{equation}
Using the definition of the Eqn.{\color{blue}(13)}, the derivative of the potential  has the form
 \begin{align}
 \begin{split}
V'(\phi)&=-\frac{-\phi}{3}\Bigg.\Bigg(\frac{\mu^{2n}(1-e^{\phi})}{(1-n)n\mu^{2}}\Bigg.\Bigg)^{\frac{1}{n-1}}+\frac{2}{3}\Bigg.\Bigg(\frac{\mu^{2n}(1-e^{\phi})}{(1-n)n\mu^{2}}\Bigg.\Bigg)^{\frac{1}{n-1}}-\frac{2(1-n)\mu^{2}}{3\mu^{2n}}\Bigg.\Bigg(\frac{\mu^{2n}(1-e^{\phi})}{(1-n)n\mu^{2}}\Bigg.\Bigg)^{\frac{n}{n-1}}.
\end{split}
\end{align}
The second derivative of Eqn.{\color{blue}(66)} gives
\begin{align}
\begin{split}
V''(\phi)&=-\frac{1}{3}\Bigg.\Bigg(\frac{\mu^{2n}(1-e^{\phi})}{(1-n)n\mu^{2}}\Bigg.\Bigg)^{\frac{1}{n-1}}+\frac{\phi e^{\phi}}{3(n-1)(1-e^{\phi})}\Bigg.\Bigg(\frac{\mu^{2n}(1-e^{\phi})}{(1-n)n\mu^{2}}\Bigg.\Bigg)^{\frac{1}{n-1}}\\
&+\frac{2(1-n)n\mu^{2}e^{\phi}}{3(n-1)\mu^{2n}(1-e^{\phi})}\Bigg.\Bigg(\frac{\mu^{2}(1-e^{\phi})}{(1-n)n\mu^{2n}}\Bigg.\Bigg)^{\frac{n}{n-1}}.
\end{split}
\end{align}
The potential is given by integration of Eqn.{\color{blue}(66)}
\begin{align}
\begin{split}
V(\phi)&=\int \frac{(2-\phi)}{3}\Bigg.\Bigg(\frac{\mu^{2n}(1-e^{\phi})}{(1-n)n\mu^{2}}\Bigg.\Bigg)^{\frac{1}{n-1}}d\phi-\frac{2(1-n)\mu^{2}}{3\mu^{2n}} \int \Bigg.\Bigg(\frac{\mu^{2n}(1-e^{\phi})}{(1-n)\mu^{2}}\Bigg.\Bigg)^{\frac{n}{n-1}} d\phi.
\end{split}
\end{align}
Using the same definitions of Eqns.{\color{blue}(15,14)} we have the expressions of the slow-roll parameters $\eta(\phi)$ and $\varepsilon(\phi)$ as
\begin{align}
\eta(\phi)&=\frac{-\frac{1}{3}\Bigg.\Bigg(\frac{\mu^{2n}(1-e^{\phi})}{(1-n)n\mu^{2}}\Bigg.\Bigg)^{\frac{1}{n-1}}+\frac{\phi e^{\phi}}{3(n-1)(1-e^{\phi})}\Bigg.\Bigg(\frac{\mu^{2n}(1-e^{\phi})}{(1-n)n\mu^{2}}\Bigg.\Bigg)^{\frac{1}{n-1}}+\frac{2(1-n)n\mu^{2}e^{\phi}}{3(n-1)\mu^{2n}(1-e^{\phi})}\Bigg.\Bigg(\frac{\mu^{2}(1-e^{\phi})}{(1-n)n\mu^{2n}}\Bigg.\Bigg)^{\frac{n}{n-1}}}{\int \frac{(2-\phi)}{3}\Bigg.\Bigg(\frac{\mu^{2n}(1-e^{\phi})}{(1-n)n\mu^{2}}\Bigg.\Bigg)^{\frac{1}{n-1}}d\phi-\frac{2(1-n)\mu^{2}}{3\mu^{2n}} \int \Bigg.\Bigg(\frac{\mu^{2n}(1-e^{\phi})}{(1-n)\mu^{2}}\Bigg.\Bigg)^{\frac{n}{n-1}} d\phi},
\begin{split}
\end{split}
\end{align}
\begin{equation}
\varepsilon(\phi)=\frac{1}{2}\frac{\Bigg.\Bigg[-\frac{-\phi}{3}\Bigg.\Bigg(\frac{\mu^{2n}(1-e^{\phi})}{(1-n)n\mu^{2}}\Bigg.\Bigg)^{\frac{1}{n-1}}+\frac{2}{3}\Bigg.\Bigg(\frac{\mu^{2n}(1-e^{\phi})}{(1-n)n\mu^{2}}\Bigg.\Bigg)^{\frac{1}{n-1}}-\frac{2(1-n)\mu^{2}}{3\mu^{2n}}\Bigg.\Bigg(\frac{\mu^{2n}(1-e^{\phi})}{(1-n)n\mu^{2}}\Bigg.\Bigg)^{\frac{n}{n-1}}
\Bigg.\Bigg]^{2}}{\Bigg.\Bigg[\int \frac{(2-\phi)}{3}\Bigg.\Bigg(\frac{\mu^{2n}(1-e^{\phi})}{(1-n)n\mu^{2}}\Bigg.\Bigg)^{\frac{1}{n-1}}d\phi-\frac{2(1-n)\mu^{2}}{3\mu^{2n}} \int \Bigg.\Bigg(\frac{\mu^{2n}(1-e^{\phi})}{(1-n)\mu^{2}}\Bigg.\Bigg)^{\frac{n}{n-1}} d\phi
 \Bigg.\Bigg)^{2}}.
\end{equation}
 \begin{table}[ht]
\caption{Table about numerical values of spectral index $n_{s}$ and tensor-to-scalar ratio $r$ for $f(R)=R-(1-n) \mu^{2}\frac{R^{n}}{\mu^{2n}}$   using reduced Planck units.} 
\centering 
\begin{tabular}{c c c c c c c} 
\hline
$n$ & $\mu$ & $\phi$ &  $ r $ & $ n_{s}$ & $r(Planck data)$  \cite{ade2016planck}& $n_{s} (Planck data)$  \cite{ade2016planck}\\ [0.5ex] 

\hline 
 1.2& 8 & 0.4& 1.00273$\times10^{-6}$ & 0.97380  & $<$0.11 &0.968$\pm 0.006$  \\   \hline

 1.2& 8 & 0.401& 1.03641$\times10^{-6}$ & 0.97335 & $<$0.11 &0.968$\pm 0.006$  \\   \hline

 1.2& 8 & 0.403& 1.10700$\times10^{-6}$ & 0.97241 & $<$0.11 &0.968$\pm 0.006$  \\   \hline

1.2 & 8 & 0.405& 1.18213$\times10^{-6}$& 0.97144 & $<$0.11 &0.968$\pm 0.006$  \\   \hline

 1.2& 8 & 0.407& 1.26204$\times10^{-6}$ & 0.97044 & $<$0.11 &0.968$\pm 0.006$ \\    \hline

 1.2& 9 & 0.409&1.34705$\times10^{-6}$& 0.96942 & $<$0.11 &0.968$\pm 0.006$  \\   \hline

 1.2& 9 & 0.410& 1.39155$\times10^{-6}$ & 0.96889 & $<$0.11 &0.968$\pm 0.006$  \\   \hline

 1.2& 9 & 0.411&1.43745$\times10^{-6}$ & 0.96836 & $<$0.11 &0.968$\pm 0.006$  \\   \hline

1.2& 9 & 0.412& 1.48477$\times10^{-6} $& 0.96781 & $<$0.11 &0.968$\pm 0.006$  \\   \hline

 1.2& 9 & 0.413& 1.53356$\times10^{-6} $& 0.96726& $<$0.11 &0.968$\pm 0.006$ \\ [1ex]

\hline 
\end{tabular}
\label{table 6} 
\end{table}
\newpage
The $KGE$ {\color{blue}(12)} in this model is rewritten as
\begin{multline}
\ddot{\phi}+\frac{1}{3}\Bigg.\Bigg[\Bigg.\Bigg(2\Big.\Big(1-e^{\phi}\Big.\Big)^{\frac{1}{n-1}}\Big.\Big(\frac{\mu^{2n-2}}{n(1-n)}\Big.\Big)^{\frac{1}{n-1}}-\Big.\Big(\frac{\mu^{2n-2}}{n(1-n)}\Big.\Big)^{\frac{n}{n-1}}e^{\phi}\Bigg.\Bigg)\Big.\Big(1-e^{\phi}\Big.\Big)^{\frac{1}{n-1}}\\-\Big.\Big(1-n\Big.\Big)^{2-2n}\Big.\Big(\frac{\mu^{2n-2}}{n(1-n)}\Big.\Big)^{\frac{n}{n-1}}\Big.\Big(1-e^{\phi}\Big.\Big)^{\frac{n}{n-1}}\Bigg.\Bigg]=0.
\end{multline}
The numerical solution to Eqn.{\color{blue}(71)} is presented in Figure {\color{blue}6}. The behaviour  of solution  is decreasing with the increase of time.
\begin{figure}[H]
  \centering
\includegraphics[height=8cm,width=13cm]{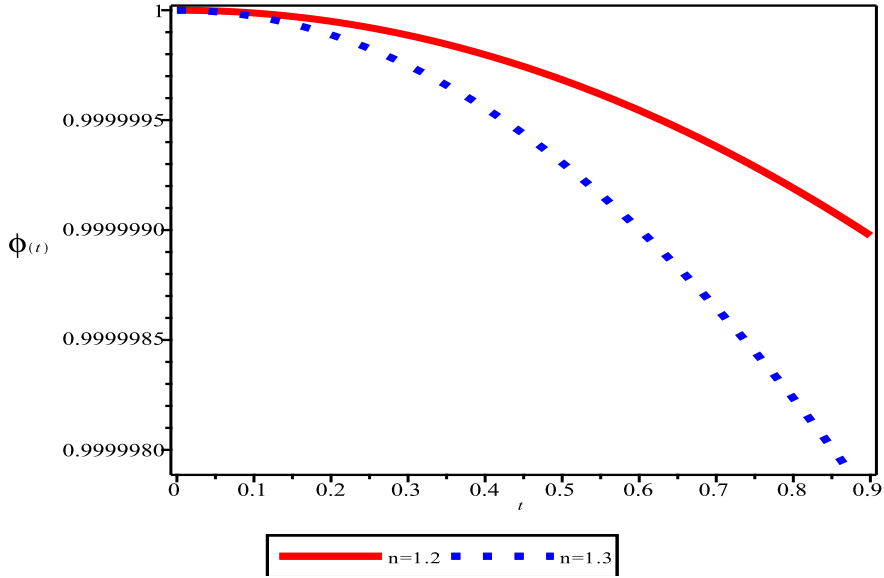}
\caption{ The scalar field $\phi(t)$ versus time $t$ for Eqn.{\color{blue}(71)}; for  $\mu=8$, $n=1.2$, $n=1.3$, $\phi(t=0)=0.405$ and $\phi'(t=0)=0$   using reduced Planck units.}
\label{fig:6}
\end{figure}
\subsection{Model7: $f(R)=R+\frac{R^{2}}{6M^{2}}$}
The case where $f(R)$ model is given by \cite{de2010f}
\begin{equation}
f(R)=R+\frac{R^{2}}{6M^{2}}
\end{equation} 
This model is known as one  of the Starobinsky models. Using the definition of the scalar field, the function $f(R)$ in terms of scalar field has the following form
\begin{equation}
f(\phi)=3M^{2}\Big.\Big(e^{\phi}-1\Big.\Big)+\frac{3}{2}M^{2}\Big.\Big(e^{\phi}-1\Big.\Big)^{2}.
\end{equation}
The effective potential $V'(\phi)$ of this model is defined as
\begin{equation}
V'{(\phi)} = -\phi M^{2} \Big.\Big(e^{\phi}-1\Big.\Big)+2M^{2}\Big.\Big(e^{\phi}-1\Big.\Big)+M^{2}\Big.\Big(e^{\phi}-1\Big.\Big)^{2}.
\end{equation}
The integration of Eqn.{\color{blue}(74)} give us the potential
\begin{equation}
V(\phi) =-M^{2}e^{\phi} \Big.\Big(\phi -\frac{3}{2} \Big.\Big)-2M^{2}\phi+\frac{M^{2} e^{2\phi}}{2}+M^{2}ln(e^{\phi})+D_{4}.
\end{equation}
The second derivative with respect to $\phi$ of Eqn.{\color{blue}(74)}
\begin{equation}
V''(\phi) =-M^{2}\Big.\Big(e^{\phi}-1\Big.\Big)-M^{2}\phi e^{\phi}+2M^{2}e^{2\phi}.
\end{equation}
Using the same definition of Eqn.{\color{blue}(14)} , we have $\varepsilon(\phi)$ parameter as
\begin{equation}
\varepsilon(\phi) = \frac{1}{2}\Bigg.\Bigg(\frac{-\phi  \Big.\Big(e^{\phi}-1\Big.\Big)+2\Big.\Big(e^{\phi}-1\Big.\Big)+\Big.\Big(e^{\phi}-1\Big.\Big)^{2}}{- e^{\phi} \Big.\Big(\phi-\frac{3}{2}\Big.\Big)-2\phi+\frac{e^{2\phi}}{2}+ln(e^{\phi})+D_{4}
}\Bigg.\Bigg)^{2}.
\end{equation}
Using the definition given by Eqn.{\color{blue}(15)}, we have $\eta(\phi)$ parameter as
\begin{equation}
\eta(\phi)=\frac{-\Big.\Big(e^{\phi}-1\Big.\Big)-\phi e^{\phi}+2 e^{2\phi}}{-e^{\phi} \Big.\Big(\phi -\frac{3}{2}\Big.\Big)-2\phi+\frac{e^{2\phi}}{2}+ln(e^{\phi})+D_{4}}.
\end{equation}
\begin{table}[!h]
\caption{Table about numerical values of spectral index $n_{s}$ and tensor-to-scalar ratio $r$ for  $f(R)=R+\frac{R^{2}}{6M^{2}}$   using reduced Planck units.} 
\centering 
\begin{tabular}{c c c c c c} 
\hline
  $D_{4}$ & $\phi$ &  $ r $ & $ n_{s}$ & $r(Planck data)$ \cite{ade2016planck} & $n_{s}(Planck data)$ \cite{ade2016planck} \\ [0.5ex] 

\hline 
 -185  & 0.25 & 0.00007 & 0.97051 & $<$0.11 &0.968$\pm 0.006$ \\    \hline

 -185  &  0.3 & 0.00012 &0.96833 & $<$0.11 &0.968$\pm 0.006$  \\   \hline

 -185  & 0.31 & 0.00013 & 0.96786 & $<$0.11 &0.968$\pm 0.006$  \\   \hline

  -185 & 0.32 & 0.00014 & 0.96738 & $<$0.11 & 0.968$\pm 0.006$ \\   \hline

 -185  & 0.35 & 0.00018 & 0.96587 & $<$0.11 &0.968$\pm 0.006$  \\    \hline

 -200  & 0.25 & 0.00006 & 0.97275 & $<$0.11 & 0.968$\pm 0.006$ \\    \hline

 -200 & 0.3 & 0.00010  & 0.97074& $<$0.11 & 0.968$\pm 0.006$  \\   \hline

 -200 & 0.31 & 0.00011  & 0.97030 & $<$0.11 & 0.968$\pm 0.006$ \\   \hline

 -200 & 0.32 & 0.00012 & 0.96986 & $<$0.11 & 0.968$\pm 0.006$ \\   \hline

-200  & 0.35 & 0.00015 & 0.96846 & $<$0.11 & 0.968$\pm 0.006$ \\    \hline

-210  & 0.25 & 0.00006 & 0.974065& $<$0.11 & 0.968$\pm 0.006$ \\    \hline

-210 & 0.3 &0.00010  & 0.97173& $<$0.11 &0.968$\pm 0.006$ \\   \hline

 -210  & 0.31 & 0.00011 & 0.97131 & $<$0. 11 &0.968$\pm 0.006$ \\   \hline

 -210  & 0.32 & 0.00013 & 0.96998 & $<$0.11 &0.968$\pm 0.006$ \\   \hline

 -210  & 0.35 & 0.00013 & 0.96998 & $<$0.11 & 0.968$\pm 0.006$ \\[1ex]

\hline 
\end{tabular}
\label{table 7} 
\end{table}
\newpage
In this model, we write the $KGE$ {\color{blue}(12)} as
\begin{equation}
\ddot{\phi}+M\Big.\Big(e^{\phi}-1\Big.\Big)\Big.\Big(3-2e^{\phi}\Big.\Big)=0
\end{equation}
The numerical solution to Eqn.{\color{blue}(79)} is presented in Figure {\color{blue}7}.
\begin{figure}[H]
  \centering
\includegraphics[height=8cm,width=13cm]{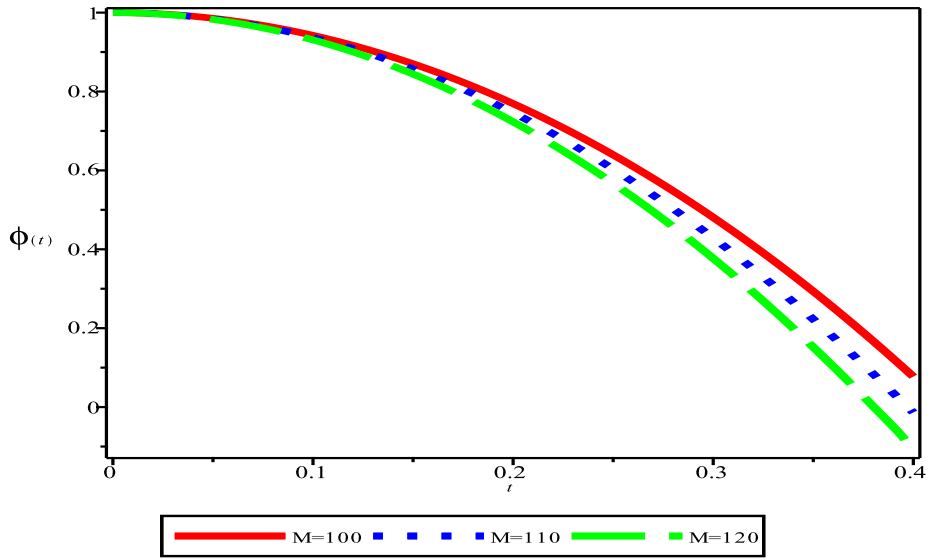}
\caption{The numerical solution of Eqn.{\color{blue}(79)}   using reduced Planck units is plotted in this figure. we used  $M=100$, $M=110$, $M=120$, $\phi(t=0)=0.32$ and $\phi'(t=0)=0$.}
\label{fig:7}
\end{figure}
\newpage
\section{Discussions}
In this study, the equivalence between $f(R)$ and scalar-tensor theories of gravity has been reviewed. We have considered seven different $f(R)$ models, namely $\beta R^{n}$,  $\alpha R+\beta R^{n}$, $R+\frac{\mu^{2n+2}}{(-R)}^{n}$, $\alpha e^{\lambda R}$,  $R-\frac{\mu^{4}}{R}$, $R-\frac{\mu^{4}}{R}$, $R-(1-n)\mu^{2}\frac{R^{n}}{\mu^{2n}}$ and $R+\frac{R^{2}}{6M^{2}}$. We constructed the potentials  from the scalar field of each $f(R)$ model. Some models produced the expressions of the potentials explicitly while others did not. For a given potential, the slow-roll approximation was applied to each of the seven $f(R)$ models to obtain  the expressions for the tensor-to-scalar ratio $r$ and spectral index $n_{s}$ parameters. The numerical computation of  $n_{s}$ and  $r$ parameters has been performed for both the exponential and polynomial models in this study. The values of $n_{s}$ and $r$ in all models were computed from Eqns.{\color{blue} (18)} and {\color{blue}(19)} respectively . The observational values of $n_{s}$ and $r$ were taken from the Planck data \cite{ade2016planck}. Observationally, the values of spectral index $n_{s}$ are in the range $n_{s}=0.968\pm 0.006$ and tensor-to-scalar ratio is the range of $r<0.11$ \cite{spergel2003first,ade2015joint,ade2016planck}. The Table {\color{blue}1-7} summarise the imposed values  of free parameters  in  such a way that the values $n_{s}$ and $r$ are consistent with the observational data (Planck data). It was observed that, scalar field $\phi$, is one of the major parameters that govern the dynamics of $n_{s}$ and $r$. 

For $f(R)=\beta R^{n}$ model, the numerical values for spectral index $n_{s}$ and tensor-to-scalar ratio $r$ are presented in Table {\color{blue}1}. One can notice that the numerical values of tensor-to-scalar ratio $r$ are consistent with Planck data results since for every fixed values of free parameter, $r$  falls below $ 0.11$ \cite{spergel2007three} and  the values of $n_{s}$  are close to the Planck data \cite{bamba2014inflationary}.  In addition, the values of $\beta$, $\phi$ and $n$ were taken to be positive constants. For specific choices of $\phi$, $n$ and $\beta$  in Table {\color{blue}1},  one can see that if $n=0.5$ and $\phi=3.2$, the obtained values of $n_{s}$ are in agreement with Planck data. For the case of $f(R)=\alpha R+\beta R^{n}$ model, the values of  spectral index $n_{s}$ and tensor-to-scalar ratio $r$ of this model have been computed and presented  in Table {\color{blue}(2)}. The results found for this model indicate  that the values of $n_{s}$  are closer to the Plank's data and those for scalar-tensor-ratio $r$ falls below $ 0.11$ as expected. Here, the values of the free parameters $\beta $ and $n$ were hold unchanged, $\alpha$  and $\phi$ were varying in both Eqns. {\color{blue}(33)} and {\color{blue} (34)}. For $f(R)=R+\frac{\mu^{2n+2}}{(-R)}^{n}$ model, numerical values of spectral index $n_{s}$ and tensor-to-scalar ratio $r$  of this model are presented in Table {\color{blue}3}. According to the numerical values of  $n_{s}$ and $r$ shown in Table  {\color{blue}3}, it is shown that $n_{s}$ are very close to the Plank's data. The numerical values of $r$ produced to this model, some fall below $ 0.11$ and others are  greater than $0.11$. For $f(R)=\alpha e^{\lambda R}$ model, the numerical values of spectral index $n_{s}$ and tensor-to-scalar ratio $r$ of this model are presented in Table {\color{blue}4}. Having a look on  these results, the values of $n_{s}$ are quite consistent with the Plank's observational data and the  values of $r$ are comparable to the Plank's data since they are $< 0.11$. The only fixed free parameter in this model is $\alpha$, others are changing. For $f(R)=R-\frac{\mu^{4}}{R}$ model, the numerical values $n_{s}$ and $r$ are listed in Table {\color{blue}5}. From there, one can see that the values of $n_{s}$ and $r$ are very close to the  Planck data. In $f(R)=R-(1-n)\mu^{2}\frac{R^{n}}{\mu^{2n}}$ model, the computational outcome of spectral index $n_{s}$ and tensor-to-scalar ratio $r$ of this model is shown in Table {\color{blue}6}. It appears that the values of $n_{s}$ and $r$ are in line with Planck data and for this case, all the parameters are changing. For $f(R)=R+\frac{R^{2}}{6M^{2}}$ model, the numerical values of $n_{s}$ and $r$ for this model are presented in Table {\color{blue}7}. It is clear that the values of $n_{s}$ and $r$ are consistent with Planck data. Similar studies of computing inflationary parameters for different types of scalar field have been obtained in \cite{ntahompagaze2017f}, \cite{ntahompagaze2018inflation}.

The $KGE$ for  each of the $f(R)$ models were also obtained. The evolution of the scalar field was studied by solving numerically different $KGE$ to seven $f(R)$ models. The numerical solutions of $KGE$ were obtained setting different values of free parameters for each $f(R)$ model. The free parameters used are the ones we have in the tables. We solved the $KGE$ setting the initial values to be  $\dot{\phi}(t_{0})=0$ and $\phi(t_{0})$   to be $\phi's$ we have in each respective table.
The scalar field $\phi(t)$ as function of time in Eqn.{\color{blue}(27)} for the case of $\beta R^{n}$ model for different values of $n$ is presented in Fig.{\color{blue}1}. The red line corresponds to the case $n=0.5$, the blue line corresponds to the case $n=0.51$ and the green one corresponds to $n=0.52$. From Fig.{\color{blue}1} one can see the decay of the scalar field as time increases. These results are similar to those found in the work done in \cite{ntahompagaze2017f} for $\alpha R^{n}$ model. 
For the case of  $\alpha R+\beta R^{n}$, we considered $n=1.5$, $n=1.6$  $n=1.7$,$\beta=0.002$ and $\alpha=0.001$, the numerical solution of Eqn.{\color{blue}(35)} is plotted in Fig.{\color{blue}2}.
The numerical computation of KGE {\color{blue}(44)} of $R+\frac{\mu^{2n+2}}{(-R)}^{n}$ is given by Fig.{\color{blue}3}, we considered $\mu=1$, $n=1.066$,$n=1.067$ and $n=1.068$.
For the case of $\alpha e^{\lambda R}$ model, we considered  $\alpha=0.002$, $\lambda=1.02$, $\lambda=1.05$ and $\lambda=1.1$, the numerical solution of Eqn.{\color{blue}(55)} is plotted in Fig.{\color{blue}4}.
For the case of  $R-\frac{\mu^{4}}{R}$ model, the numerical solution of Eqn.{\color{blue}(63)} is presented in Fig.{\color{blue}5}, we set the value of the free parameter $\mu=0.7$, $\mu=0.8$ and $\mu=0.9$.
For the case of $ R-(1-n)\mu^{2}\frac{R^{n}}{\mu^{2n}}$, we considered $n=1.2$ and $n=1.3$, the numerical solution of Eqn.{\color{blue}(71)} is plotted in Fig.{\color{blue}(6)}.
For the case of $R+\frac{R^{2}}{6M^{2}}$ , we considered $M=100$, $M=110$ and $M=120$, the numerical solution of Eqn.{\color{blue}(79)} is plotted in Fig.{\color{blue}(7)}. From  figure  {\color{blue}1-7}, the same behaviour was manifested which is the  decay of the scalar field as time increases.
\section{Conclusions }\label{conclusion1}
In this paper, the relationship between $f(R)$ and scalar-tensor theories has been reviewed. In this line, seven $f(R)$ models have been considered, focusing on the inflationary era. Using a specific definition of scalar field as the extra degree of freedom, we constructed the potential for each $f(R)$ model. We applied slow-roll approximation for each $f(R)$ model and obtained the expression of the $\varepsilon(\phi) $ and $\eta(\phi)$ parameters. We numerically computed the two inflationary parameters (spectral index $n_{s}$ and tensor-to-scalar ratio $r$) for each $f(R)$ model. We used different ranges of free parameters so that we can get the values of $n_{s}$ and  $r$ that are in the same range as the obtained values from observations. The values of $n_{s}$ and $r$ produced for all $f(R)$ models, some are in the good range  and others are very close to the Planck survey data. For the first model, the overall average values of $n_{s}$ and $r$ generated are  $0.96697$ and  $0.00155$, respectively. For the second model, we have generated $n_{s}=0.97105$ and tensor-to-scalar ratio $r=0.00149$.  For the third model, the overall average of $n_{s}$ and $r$ are respectively given by $0.97019$ and $0.07008$. For the fourth model, we have generated  the values of $n_{s}$ and $r$\ to be $0.96516$ and $0.00337$, respectively. For the fifth model, the overall average of spectral index $n_{s}$ and tensor-to-scalar ratio $r$ values are  $0.96940$ and $0.01297$, respectively. For the sixth model, the values of $n_{s}$ and $r$ are, respectively, given by $0.97032$ and $1.27846\times10^{-6}$. The seventh model, the overall average of $n_{s}$ and $r$ are $0.96994$ and $0.00011$, respectively. We also derived  the $KGE$  for each toy of  $f(R)$ model, therefore checked for the evolution of the scalar field with respect to time.  In figure {\color{blue}1}, the scalar field $\phi$ decays with the increase of time, the same behaviour  is manifested in figures {\color{blue}1-7}. The results show that there is a less content of scalar field in the late Universe. In addition to that, one can constrain the parameter space from the $f(R)$ gravity models based on the current and future cosmological observations.
\section*{Acknowledgements}
We thank the anonymous reviewers for their constructive comments towards the significant improvement of the
present manuscript. BM, AM and JN are thankful to the financial support provided by the Swedish International Development Cooperation Agency (SIDA) through the International Science Programme (ISP) to the University of Rwanda via Rwanda Astrophysics, Space and Climate Science Research Group (RASCSRG) grant number RWA01. FT and AA acknowledge the physics Department to accord them time and place to conduct this research.


\begin{thebibliography}{10}

\bibitem{guth1981cosmological}
Alan~H Guth and Erick~J Weinberg.
\newblock Cosmological consequences of a first-order phase transition in the s
  u 5 grand unified model.
\newblock {\em Physical Review D}, 23(4):876, 1981.

\bibitem{linde2008inflationary}
Andrei Linde.
\newblock Inflationary cosmology.
\newblock In {\em Inflationary Cosmology}, pages 1--54. Springer, 2008.

\bibitem{ade2016planck}
PAR Ade, N~Aghanim, M~Arnaud, F~Arroja, Mark Ashdown, J~Aumont, C~Baccigalupi,
  M~Ballardini, AJ~Banday, RB~Barreiro, et~al.
\newblock Planck 2015 results-xx. constraints on inflation.
\newblock {\em Astronomy \& Astrophysics}, 594:A20, 2016.

\bibitem{martin2016have}
J{\'e}r{\^o}me Martin.
\newblock What have the planck data taught us about inflation?
\newblock {\em Classical and Quantum Gravity}, 33(3):034001, 2016.

\bibitem{gorbunov2011introduction}
Dmitry~S Gorbunov and Valery~A Rubakov.
\newblock {\em Introduction to the theory of the early universe: Cosmological
  perturbations and inflationary theory}.
\newblock World Scientific, 2011.

\bibitem{nojiri2007econf}
S~Nojiri.
\newblock econf c 0602061 (2006) 06; s. nojiri, sd odintsov.
\newblock {\em Int. J. Geom. Meth. Mod. Phys}, 4:115, 2007.

\bibitem{capozziello2011extended}
Salvatore Capozziello and Mariafelicia De~Laurentis.
\newblock Extended theories of gravity.
\newblock {\em Physics Reports}, 509(4-5):167--321, 2011.

\bibitem{nojiri2017modified}
Sh~Nojiri, SD~Odintsov, and VK~Oikonomou.
\newblock Modified gravity theories on a nutshell: inflation, bounce and
  late-time evolution.
\newblock {\em Physics Reports}, 692:1--104, 2017.

\bibitem{odintsov2016inflationary}
SD~Odintsov and VK~Oikonomou.
\newblock Inflationary $\alpha$-attractors from ${F (R)}$ gravity.
\newblock {\em Physical Review D}, 94(12):124026, 2016.

\bibitem{odintsov2017inverse}
SD~Odintsov and VK~Oikonomou.
\newblock Inverse symmetric inflationary attractors.
\newblock {\em Classical and Quantum Gravity}, 34(10):105009, 2017.

\bibitem{riess1998observational}
Adam~G Riess, Alexei~V Filippenko, Peter Challis, Alejandro Clocchiatti, Alan
  Diercks, Peter~M Garnavich, Ron~L Gilliland, Craig~J Hogan, Saurabh Jha,
  Robert~P Kirshner, et~al.
\newblock Observational evidence from supernovae for an accelerating universe
  and a cosmological constant.
\newblock {\em The Astronomical Journal}, 116(3):1009, 1998.

\bibitem{perlmutter1999constraining}
Saul Perlmutter, Michael~S Turner, and Martin White.
\newblock Constraining dark energy with type ia supernovae and large-scale
  structure.
\newblock {\em Physical Review Letters}, 83(4):670, 1999.

\bibitem{capozziello2002curvature}
Salvatore Capozziello.
\newblock Curvature quintessence.
\newblock {\em International Journal of Modern Physics D}, 11(04):483--491,
  2002.

\bibitem{carloni2005cosmological}
Sante Carloni, Peter~KS Dunsby, Salvatore Capozziello, and Antonio Troisi.
\newblock Cosmological dynamics of ${Rn}$ gravity.
\newblock {\em Classical and Quantum Gravity}, 22(22):4839, 2005.

\bibitem{cembranos2006dark}
JAR Cembranos, A~Dobado, and AL~Maroto.
\newblock Dark matter clues in the muon anomalous magnetic moment.
\newblock {\em Physical Review D}, 73(5):057303, 2006.

\bibitem{nojiri2006dark}
Shin’ichi Nojiri, Sergei~D Odintsov, and M~Sami.
\newblock Dark energy cosmology from higher-order, string-inspired gravity, and
  its reconstruction.
\newblock {\em Physical Review D}, 74(4):046004, 2006.

\bibitem{sotiriou2006nearly}
Thomas~P Sotiriou.
\newblock The nearly newtonian regime in non-linear theories of gravity.
\newblock {\em General Relativity and Gravitation}, 38(9):1407--1417, 2006.

\bibitem{sawicki2007stability}
Ignacy Sawicki and Wayne Hu.
\newblock Stability of cosmological solutions in ${f (R)}$ models of gravity.
\newblock {\em Physical Review D}, 75(12):127502, 2007.

\bibitem{carloni2010covariant}
Sante Carloni.
\newblock Covariant gauge invariant theory of scalar perturbations in ${f
  (R)}$-gravity: A brief review.
\newblock {\em The Open Astronomy Journal}, 3(2):76--93, 2010.

\bibitem{sotiriou2006f}
Thomas~P Sotiriou.
\newblock ${f (R)}$ gravity and scalar--tensor theory.
\newblock {\em Classical and Quantum Gravity}, 23(17):5117, 2006.

\bibitem{nojiri2003modified}
Shin’ichi Nojiri and Sergei~D Odintsov.
\newblock Modified gravity with negative and positive powers of curvature:
  Unification of inflation and cosmic acceleration.
\newblock {\em physical Review D}, 68(12):123512, 2003.

\bibitem{odintsov2021r}
Odintsov, Sergei D and Oikonomou, VK.
\newblock R 2 inflation revisited and dark energy corrections.
\newblock {\em Physical Review D}, 104(12):124065, 2021.

\bibitem{appleby2008aspects}
Appleby, Stephen A and Battye, Richard A.
\newblock Aspects of cosmological expansion in F (R) gravity models.
\newblock {\em Journal of Cosmology and Astroparticle Physics}, 2008(05):019, 2008.

\bibitem{oikonomou2021rescaled}
Oikonomou, VK.
\newblock Rescaled Einstein-Hilbert gravity from f (R) gravity: Inflation, dark energy, and the swampland criteria.
\newblock {\em Physical Review D}, 103(12):124028, 2021

\bibitem{sepehri2015unifying}
Sepehri, Alireza and Rahaman, Farook and Setare, Mohammad Reza and Pradhan, Anirudh and Capozziello, Salvatore and Sardar, Iftikar Hossain.
\newblock Unifying inflation with late-time acceleration by a BIonic system.
\newblock {\em Physics Letters B}, 747(1--8): 2015

\bibitem{yashiki2020local}
Yashiki, Mai and Sakai, Nobuyuki and Saito, Ryo.
\newblock Local-gravity test of unified models of inflation and dark energy in f (R) gravity.
\newblock {\em Physical Review D}, 102(4): 043504,2020

\bibitem{capozziello2000nonminimal}
Capozziello, S and Lambiase, G and Schmidt, H-J.
\newblock Nonminimal derivative couplings and inflation in generalized theories of gravity.
\newblock {\em Annalen der Physik}, 9(1): 39--48,2000

\bibitem{oikonomou2021unifying}
Oikonomou, VK.
\newblock Unifying inflation with early and late dark energy epochs in axion F (R) gravity.
\newblock {\em Physical Review D}, 103(4): 044036,2021

\bibitem{capozziello2006unified}
Capozziello, S and Nojiri, S’i and Odintsov, SD.
\newblock Unified phantom cosmology: Inflation, dark energy and dark matter under the same standard.
\newblock {\em Physics Letters B}, 632(5-6): 597--604,2006

\bibitem{odintsov2020aspects}
Odintsov, Sergei D and Oikonomou, VK.
\newblock Aspects of axion F (R) gravity.
\newblock {\em EPL (Europhysics Letters)}, 129(4):40001,2020

\bibitem{odintsov2020geometric}
Odintsov, Sergei D and Oikonomou, VK.
\newblock Geometric inflation and dark energy with axion F (R) gravity.
\newblock {\em Physical Review D}, 101(4):044009,2020

\bibitem{de2011generalizing}
De-Santiago, Josue and Cervantes-Cota, Jorge L.
\newblock Generalizing a unified model of dark matter, dark energy, and inflation with a noncanonical kinetic term.
\newblock {\em Physical Review D}, 83(6):063502,2011

\bibitem{odintsov2019unification}
Odintsov, SD and Oikonomou, VK.
\newblock Unification of inflation with dark energy in f (R) gravity and axion dark matter.
\newblock {\em Physical Review D}, 99(10):104070,2019

\bibitem{starobinsky1980new}
Alexei~A Starobinsky.
\newblock A new type of isotropic cosmological models without singularity.
\newblock {\em Physics Letters B}, 91(1):99--102, 1980.

\bibitem{nojiri2007unifying}
Shin'ichi Nojiri and Sergei~D Odintsov.
\newblock Unifying inflation with ${\Lambda}$cdm epoch in modified ${f (R)}$
  gravity consistent with solar system tests.
\newblock {\em Physics Letters B}, 657(4-5):238--245, 2007.

\bibitem{nojiri2008future}
Shin’ichi Nojiri and Sergei~D Odintsov.
\newblock Future evolution and finite-time singularities in ${f (R)}$ gravity
  unifying inflation and cosmic acceleration.
\newblock {\em Physical Review D}, 78(4):046006, 2008.

\bibitem{barrow2018reconstructions}
John~D Barrow and Andronikos Paliathanasis.
\newblock Reconstructions of the dark-energy equation of state and the
  inflationary potential.
\newblock {\em General relativity and gravitation}, 50(7):1--25, 2018.

\bibitem{ntahompagaze2017f}
Joseph Ntahompagaze, Amare Abebe, and Manasse Mbonye.
\newblock On ${f (R)}$ gravity in scalar--tensor theories.
\newblock {\em International Journal of Geometric Methods in Modern Physics},
  14(07):1750107, 2017.

\bibitem{ntahompagaze2018inflation}
Joseph Ntahompagaze, Jean Damasc{\`e}ne~Mbarubucyeye, Shambel Sahlu, and Amare
  Abebe.
\newblock Inflation constraints for classes of ${f (R)}$ models.
\newblock {\em International Journal of Geometric Methods in Modern Physics},
  15(12):1850209, 2018.

\bibitem{sebastiani2014nearly}
L~Sebastiani, G~Cognola, R~Myrzakulov, SD~Odintsov, and S~Zerbini.
\newblock Nearly starobinsky inflation from modified gravity.
\newblock {\em Physical Review D}, 89(2):023518, 2014.

\bibitem{oikonomou2021refined}
Oikonomou, VK.
\newblock A refined Einstein--Gauss--Bonnet inflationary theoretical framework.
\newblock {\em Classical and Quantum Gravity}, 38(19):195025,2021

\bibitem{myrzakulov2016inflationary}
Ratbay Myrzakulov, SD~Odintsov, and Lorenzo Sebastiani.
\newblock Inflationary universe from higher derivative quantum gravity coupled
  with scalar electrodynamics.
\newblock {\em Nuclear Physics B}, 907:646--663, 2016.

\bibitem{oikonomou2015singular}
Oikonomou, VK.
\newblock Singular bouncing cosmology from Gauss-Bonnet modified gravity.
\newblock {\em Physical Review D},  92(12):124027, 2015.

\bibitem{odintsov2017unification}
SD~Odintsov, VK~Oikonomou, and L~Sebastiani.
\newblock Unification of constant-roll inflation and dark energy with
  logarithmic ${R2}$-corrected and exponential ${f (R)}$ gravity.
\newblock {\em Nuclear Physics B}, 923:608--632, 2017.

\bibitem{murorunkwere20211+}
Beatrice Murorunkwere, Joseph Ntahompagaze, and Edward Jurua.
\newblock 1+ 3 covariant perturbations in power-law ${f (R)}$ gravity.
\newblock {\em The European Physical Journal C}, 81(4):1--10, 2021.

\bibitem{sotiriou2010f}
Thomas~P Sotiriou and Valerio Faraoni.
\newblock ${f (R)}$ theories of gravity.
\newblock {\em Reviews of Modern Physics}, 82(1):451, 2010.

\bibitem{clifton2012modified}
Timothy Clifton, Pedro~G Ferreira, Antonio Padilla, and Constantinos Skordis.
\newblock Modified gravity and cosmology.
\newblock {\em Physics reports}, 513(1-3):1--189, 2012.

\bibitem{frolov2008singularity}
Andrei~V Frolov.
\newblock Singularity problem with ${f (R)}$ models for dark energy.
\newblock {\em Physical review letters}, 101(6):061103, 2008.

\bibitem{odintsov2017inflation}
SD~Odintsov and VK~Oikonomou.
\newblock Inflation with a smooth constant-roll to constant-roll era
  transition.
\newblock {\em Physical Review D}, 96(2):024029, 2017.

\bibitem{oikonomou2017reheating}
VK~Oikonomou.
\newblock Reheating in constant-roll ${f(R)}$ gravity.
\newblock {\em Modern Physics Letters A}, 32(33):1750172, 2017.

\bibitem{copeland1993reconstructing}
Edmund~J Copeland, Edward~W Kolb, Andrew~R Liddle, and James~E Lidsey.
\newblock Reconstructing the inflaton potential: In principle and in practice.
\newblock {\em Physical Review D}, 48(6):2529, 1993.

\bibitem{liddle1992cobe}
Andrew~R Liddle and David~H Lyth.
\newblock Cobe, gravitational waves, inflation and extended inflation.
\newblock {\em Physics Letters B}, 291(4):391--398, 1992.

\bibitem{komatsu2002pursuit}
Eiichiro Komatsu.
\newblock {\em The Pursuit of Non-Gaussian Fluctuations in the Cosmic Microwave
  Background}.
\newblock PhD thesis, Citeseer, 2002.

\bibitem{liddle1994formalizing}
Andrew~R Liddle, Paul Parsons, and John~D Barrow.
\newblock Formalizing the slow-roll approximation in inflation.
\newblock {\em Physical Review D}, 50(12):7222, 1994.

\bibitem{bassett2006inflation}
Bruce~A Bassett, Shinji Tsujikawa, and David Wands.
\newblock Inflation dynamics and reheating.
\newblock {\em Reviews of Modern Physics}, 78(2):537, 2006.

\bibitem{huang2014polynomial}
Qing-Guo Huang.
\newblock A polynomial ${f (R)}$ inflation model.
\newblock {\em Journal of Cosmology and Astroparticle Physics}, 2014(02):035,
  2014.

\bibitem{barrow1983stability}
John~D Barrow and Adrian~C Ottewill.
\newblock The stability of general relativistic cosmological theory.
\newblock {\em Journal of Physics A: Mathematical and General}, 16(12):2757,
  1983.

\bibitem{carloni2006bounce}
Sante Carloni, Peter~KS Dunsby, and Deon Solomons.
\newblock Bounce conditions in ${f (R)}$ cosmologies.
\newblock {\em Classical and Quantum Gravity}, 23(6):1913, 2006.

\bibitem{munyeshyaka2021cosmological}
Munyeshyaka Albert, Ntahompagaze Joseph, and Mutabazi Tom.
\newblock Cosmological perturbations in ${f (G)}$ gravity.
\newblock {\em International Journal of Modern Physics D},
  30(07):2150053, 2021.
  
  \bibitem{li2007cosmology}
Baojiu Li and John~D Barrow.
\newblock Cosmology of ${f (R)}$ gravity in the metric variational approach.
\newblock {\em Physical Review D}, 75(8):084010, 2007.

\bibitem{faulkner2007constraining}
Thomas Faulkner, Max Tegmark, Emory~F Bunn, and Yi~Mao.
\newblock Constraining ${f (R)}$ gravity as a scalar-tensor theory.
\newblock {\em Physical Review D}, 76(6):063505, 2007.

\bibitem{carroll2004cosmic}
Sean~M Carroll, Vikram Duvvuri, Mark Trodden, and Michael~S Turner.
\newblock Is cosmic speed-up due to new gravitational physics?
\newblock {\em Physical Review D}, 70(4):043528, 2004.

\bibitem{de2010f}
Antonio De~Felice and Shinji Tsujikawa.
\newblock ${f (R)}$ theories.
\newblock {\em Living Reviews in Relativity}, 13(1):1--161, 2010.

\bibitem{spergel2003first}
David~N Spergel, Licia Verde, Hiranya~V Peiris, Eiichiro Komatsu, MR~Nolta,
  Charles~L Bennett, Mark Halpern, Gary Hinshaw, Norman Jarosik, Alan Kogut,
  et~al.
\newblock First-year wilkinson microwave anisotropy probe (wmap)* observations:
  determination of cosmological parameters.
\newblock {\em The Astrophysical Journal Supplement Series}, 148(1):175, 2003.

\bibitem{ade2015joint}
Peter~AR Ade, N~Aghanim, Z~Ahmed, RW~Aikin, Kate~Denham Alexander, M~Arnaud,
  J~Aumont, C~Baccigalupi, Anthony~J Banday, D~Barkats, et~al.
\newblock Joint analysis of bicep2/keck array and planck data.
\newblock {\em Physical review letters}, 114(10):101301, 2015.

\bibitem{spergel2007three}
David~N Spergel, R~Bean, O~Dor{\'e}, MR~Nolta, CL~Bennett, Joanna Dunkley,
  G~Hinshaw, N~ea Jarosik, E~Komatsu, L~Page, et~al.
\newblock Three-year wilkinson microwave anisotropy probe (wmap) observations:
  implications for cosmology.
\newblock {\em The Astrophysical Journal Supplement Series}, 170(2):377, 2007.

\bibitem{bamba2014inflationary}
Kazuharu Bamba, Shin’ichi Nojiri, Sergei~D Odintsov, and Diego Saez-Gomez.
\newblock Inflationary universe from perfect fluid and ${f (R)}$ gravity and
  its comparison with observational data.
\newblock {\em Physical Review D}, 90(12):124061, 2014.



\end{thebibliography}
\end{document}